\documentclass{article}
\usepackage[12pt]{extsizes}
\usepackage[utf8]{inputenc}
\usepackage{amsmath}
\usepackage{amssymb}
\usepackage{amsthm}
\usepackage{natbib}
\usepackage{graphicx}
\usepackage{lipsum}
\usepackage[dvipsnames]{xcolor}
\usepackage[hidelinks]{hyperref}
\hypersetup{
    colorlinks = true,
    linkcolor={blue!50!black},
    citecolor={blue!50!black},
    urlcolor={blue!80!black}
}
\usepackage{authblk}
\usepackage{geometry}
\usepackage{algorithm,algorithmicx,algpseudocode}
\usepackage[p,osf]{cochineal}
\usepackage[scale=.95,type1]{cabin}
\usepackage[cochineal,bigdelims,cmintegrals,vvarbb]{newtxmath}
\usepackage[zerostyle=c,scaled=.94]{newtxtt}
\usepackage[cal=boondoxo]{mathalfa}
\usepackage{microtype}

\usepackage{etoolbox}
\apptocmd{\thebibliography}{\raggedright}{}{}

\usepackage{amsmath,amsfonts,bm}

\def\eqref#1{equation~\ref{#1}}

\def\va{{\bm{a}}}

\DeclareMathAlphabet{\mathsfit}{\encodingdefault}{\sfdefault}{m}{sl}
\SetMathAlphabet{\mathsfit}{bold}{\encodingdefault}{\sfdefault}{bx}{n}

\def\gT{{\mathcal{T}}}

\newcommand{\calI}{\mathcal{I}}

\def\sA{{\mathbb{A}}}

\newcommand{\E}{\mathbb{E}}

\newcommand{\V}[1]{\bm{#1}}

\newcommand{\eq}[1]{\begin{align}#1\end{align}}

\newcommand{\fr}[2]{\frac{#1}{#2}}
\newcommand{\brck}[1]{\left(#1\right)}
\newcommand{\brcka}[1]{\left\langle#1\right\rangle}
\newcommand{\brcksq}[1]{\left[#1\right]}
\newcommand{\brckcur}[1]{\left\{#1\right\}}

\def\weight{w}
\def\mweight{\theta}
\def\pweight{\phi}

\newcommand{\policy}{\pi}

\def\ob{o}
\def\Ac{\va}
\def\ac{a}

\def\st{s}

\def\rew{r}

\def\df{\gamma}
\def\pol{\pi}

\def\eplen{H}
\def\samplinghorizon{\mathcal{h}}
\def\trans{\gT}
\def\hidden{h}

\def\util{u}

\def\utility{\util}

\def\endow{x}
\def\vendow{\V{\endow}}

\newcommand{\posttax}[1]{\tilde{#1}}

\def\skill{n}

\def\elas{e}

\def\income{z}
\def\consumption{c}

\def\labor{l}
\def\vlabor{\V{\labor}}

\def\crra{\texttt{crra}}

\def\isoeta{\eta}

\def\tax{T}
\def\taxrate{\tau}
\def\mtaxrate{\tau}
\def\taxperiodlen{M}

\def\bracketcutoff{m}

\def\incometax{ T }  %

\def\money{\endow^c}

\def\stone{\endow^s}
\def\wood{\endow^w}

\def\Money{\vendow^c}

\newcommand{\mon}[2]{\money_{#1,#2}}

\def\pdf{f}
\def\cdf{F}
\def\gini{\texttt{gini}}
\def\equality{\texttt{eq}}
\def\productivity{\texttt{prod}}

\def\plannerpolicy{\pi_p}
\def\socialwelfare{\texttt{swf}}

\def\socialwelfareweight{\omega}
\def\paretoweight{g}
\def\socialmarginalwelfareweight{\paretoweight}
\def\camelback{"Camelback"}

\title{
The AI Economist: \\
Improving Equality and Productivity with AI-Driven Tax Policies}

\author[*,1]{Stephan Zheng}
\author[*,1]{Alexander Trott}
\author[1]{Sunil Srinivasa}
\author[1]{Nikhil Naik}
\author[1]{Melvin Gruesbeck}
\author[1,2]{David C. Parkes}
\author[1]{Richard Socher}

\affil[1]{Salesforce Research}
\affil[2]{Harvard University}

\date{April 28, 2020}

\begin{document}
\maketitle
{\let\thefootnote\relax\footnote{
* indicates significant contribution.
R.S. and S.Z. conceived and directed the project;
S.Z., A.T., N.N., and D.P. developed the theoretical framework;
A.T. and S.Z. developed the economic simulator;
A.T. and S.Z. implemented the reinforcement learning platform and performed experiments with AI agents;
A.T., S.Z., and D.P. processed and analyzed experiments with AI agents;
S.Z. implemented and performed the experiments with human participants;
M.G., N.N., and S.Z. designed the interface for human participants;
S.S. and S.Z. processed the results with human participants;
S.Z., A.T., S.S., N.N., and D.P. interpreted the results with human participants;
A.T., M.G., S.S., and S.Z. designed the figures and visualizations;
S.Z., A.T., N.N., and D.P. drafted the manuscript;
Kathy Baxter drafted the ethical review;
R.S. planned and advised the work, and analyzed all results;
all authors discussed the results and commented on the manuscript.
We thank Kathy Baxter,
Lofred Madzou, Simon Chesterman, Rob Reich, Mia de Kuijper, Scott Kominers, Gabriel Kriendler, Stefanie Stantcheva, and Thomas Piketty for valuable discussions.
This research was not conducted with any corporate or commercial applications in mind.
Correspondence to:
\url{stephan.zheng@salesforce.com}.
}}
\vspace{-.66in}
\begin{abstract}
Tackling real-world socio-economic challenges requires designing and testing economic policies.
However, this is hard in practice, due to a lack of appropriate (micro-level) economic data and limited opportunity to experiment.
In this work, we train social planners that discover tax policies in dynamic economies that can effectively trade-off economic equality and productivity.
We propose a two-level deep reinforcement learning approach to learn \emph{dynamic tax policies}, based on economic simulations in which both agents and a government learn and adapt.
Our data-driven approach does not make use of economic modeling assumptions, and learns from observational data alone.
We make four main contributions.
First, we present an economic simulation environment that features competitive pressures and market dynamics.
We validate the simulation by showing that baseline tax systems perform in a way that is consistent with economic theory, including in regard to learned agent behaviors and specializations.
Second, we show that AI-driven tax policies improve the trade-off between equality and productivity by 16\% over baseline policies, including the prominent Saez tax framework.
Third, we showcase several emergent features: AI-driven tax policies are qualitatively different from baselines, setting a higher top tax rate and higher net subsidies for low incomes.
Moreover, AI-driven tax policies perform strongly in the face of emergent tax-gaming strategies learned by AI agents.
Lastly, AI-driven tax policies are also effective when used in experiments with human participants.
In experiments conducted on MTurk, an AI tax policy provides an equality-productivity trade-off that is similar to that provided by the Saez framework along with higher inverse-income weighted social welfare.
\end{abstract}
\section{Introduction}
Economic inequality is accelerating globally and is a key social and economic concern.
Many studies have shown that large income inequality gaps can have significant negative effects,  leading for example to diminished economic opportunity \citep{inequality_matters_2013} and adverse health effects \citep{subramanian2004income}.
In this light, tax policy provides governments with an important tool to reduce inequality, supporting the possibility of the redistribution of wealth through government provided services and benefits.
And yet, finding the optimal tax policy is challenging.
The basic reason is that while more taxation can improve equality, taxation can also discourage people from working, leading to lower productivity.

The problem of optimally balancing equality and productivity has not been solved for general economic settings, and even when the policy objectives can be agreed upon.
Part of the challenge is that is hard to experiment with real-world tax policies.
In the place of experimentation, economic theory often relies on simplifying assumptions that are hard to validate, for example about people’s sensitivity to taxes.
Tax systems that have been proposed range from no taxes at all (“free market”), to progressive and regressive tax systems (reflecting whether the tax rate increases or decreases as income increases), to total redistribution.

In this paper, we introduce “The AI Economist,” a {\em two-level deep reinforcement learning} (RL) framework to train \emph{social planners}. The economic actors are adaptive, learning behaviors in the simulated world and including in response to tax policy. The planner is also adaptive, learning  tax policies that adapt to agent behaviors and seek to achieve a particular policy objective.
Neither economic actors nor the AI Economist have prior knowledge, whether about the simulated world environment or  economic theory.
The AI Economist learns a tax policy based only on observable data and without knowledge of the skill or utility functions of workers or prior assumptions about the behavior of workers, and can be used to optimize for any desired social outcome.

The AI Economist learns a {\em tax schedule}, analogous to the way in which US federal income taxes are described. Taxes are computed by applying a tax rate to each part of an individual's income that falls within a tax bracket.
For simplicity, we fix the intervals that correspond to  each of these income brackets and  learn the tax rate for each bracket.
The tax schedule learned by the AI Economist is not personalized;  each agent faces the same rates and bracket cutoffs. In a single tax period the tax schedule is determined via a deep neural network, able to observe all public information about the world, including the position, income, and resources held by agents.

Our approach to economic design is based on the use of   simulations, making use of AI agents that learn optimal behaviors.
This use of  simulation enables the testing of economic policies at large-scale, and including the ability to measure a range of different metrics.
In effect, we can  compare the performance of millions of economic designs, making use of economic agents whose behavior is learned in parallel.
The simulation framework can also be used to speed up experiments with existing proposals for tax systems, validating assumptions and offering the ability to test ideas that come from  economic theory. 

We make the following contributions:

\begin{itemize}
  \item We introduce a principled economic simulation that features competitive pressures, trade, and resource scarcity.
  \item We validate that learned  behavior conforms to results known from economic theory, for example agent specialization.
  \item We frame the problem of learning optimal taxes in a dynamic economy as a \emph{two-level, inner-outer reinforcement learning problem} and   describe a range of techniques to stabilize training for this two-level RL problem, including the use of learning curricula and entropy-based regularization.
  \item The AI-driven tax policies make use of different kinds of tax rate schedules than those suggested by baseline policies, and our experiments demonstrate that the AI-driven tax policy can improve the trade-off between equality and productivity by 16\% when compared to the prominent Saez tax framework.
  \item We show that AI agents can learn tax-avoidance behaviors, modulating their incomes across tax periods. The tax schedule generated by the AI Economist performs well despite this kind of strategic behavior.
  \item Without endorsing the particular tax schedules, we show that a learned policy can also be effective in experiments with human participants and without additional recalibration. The policy achieves an equality-productivity trade-off that is competitive with the state-of-the-art, together with higher inverse-income weighted social welfare. This provides a preliminary suggestion that the AI Economist methodology could also be applicable to more general, real world settings.
\end{itemize}

\subsection{Related Work}
\label{sec:relatedwork}
\paragraph{Optimal Taxation.}
In economics, optimal tax theory is the study of the design of a tax system that maximizes a social welfare function subject to a set of economic constraints, while accounting for the fact that individuals respond to taxes and transfers~\citep{mankiw_optimal_2009,diamond_case_2011}.
The core challenge in the design of optimal tax policies is that taxes and transfers can affect
incentives to work, creating a trade-off between equality and productivity~\citep{mankiw_optimal_2009,diamond_case_2011}.
A particular concern is that high income may correlate with high skill, leading  higher skilled workers to choose to work less.

\cite{ramsey1927contribution}'s early  work tied consumption taxes on a good to a representative consumer's elasticity of demand for the good.
The current dominant theoretical framework  arose out of a series of papers by Mirrlees and Diamond~\citep{diamond1971optimal, diamond1971optimal2,mirrlees_optimal_1976}.
These authors consider a utilitarian social planner---aiming to maximize the sum of individual utilities in a society.
\cite{saez_using_2001} builds on the Mirrlees framework to derive optimal non-linear tax rates using models of the elasticity of earnings with respect to tax rates, together with the shape of the income distribution.

Other work has expanded upon the Mirrlees framework to argue for a tax system that tries to achieve a broader distributive justice~\citep{piketty2013theory,piketty2014optimal,saez2016generalized}, or a tax system in which the payments made by an individual merely match the benefits received~\citep{mankiw_optimal_2009,mankiw2010spreading,mankiw2010optimal}.

The Mirrlees model is limited to optimal taxation in a single tax period, without considering dynamics, for example the income histories of individuals in deciding taxes, or events with longer-term effects such as education.
The {\em new dynamic public finance} (NDPF) expands upon these frameworks to consider dynamic economies, capturing additional real world effects, for example, allowing for the coordinated taxation of capital and labor income~\citep{golosov2003optimal,kocherlakota2005zero,albanesi2006dynamic,kocherlakota_new_2010}.

Progress in optimal taxation theory has also come through a growing empirical and experimental literature.
This includes work that seeks to estimate labor supply elasticity to changes in taxation and redistribution~\citep{gruber2002elasticity,chetty2012bounds,goldberg2016kwacha}, and work that seeks to understand the behavioral response of workers to tax policy through the use of cross-sectional data on taxation, labor supply, and individual incomes~\citep{slemrod1996high,goolsbee2000happens,alesina2005work}.
Research in behavioral public finance~\citep{mccaffery2006behavioral,kuziemko2015elastic,alesina2018intergenerational} makes use of experiments and surveys to understand how
people respond to different theories of taxation, redistribution, and public spending.

Our work adopts baselines from  optimal taxation theory, by comparing the performance of the AI Economist with tax policies that arise from the Saez  framework, in this case, making use of estimated labor elasticities in our simulated economies.
\paragraph{Agent-based Modeling.}

Agent-based modeling (ABM) research~\citep{holland1991artificial,bonabeau_agent-based_2002} creates simulations of agents and institutions that interact through prescribed rules.
ABM does not rely on standard equilibrium models. Rather, it allows for dynamic, nonlinear behavior by agents and institutions, and can adopt behavioral rules that are deduced from human experiments~\citep{arthur_designing_1991}.

The idea is to use ABM to enable policy-makers to simulate an artificial economy under different policy scenarios, and quantitatively explore their consequences~\citep{farmer_economy_2009}.
ABM has been applied to study tax compliance~\citep{bloomquist_tax_2011-1,miguel_exploring_2012,subburaj_theory_2018}, and to derive optimal taxation policy~\citep{garrido_agent_2013}, based on heuristics and simple learning methods.
Wider adoption of ABM has proved challenging due to the complexity of realistically modeling human behavior and the economy.

While our motivations are similar to ABM, our framework makes use of deep RL to optimize the behaviors of economic agents with the effect that we study policy design in the presence of rational agent behavior.

\paragraph{Reinforcement Learning.}

Our learning approach relates to {\em multi-agent reinforcement learning} (MARL).
In MARL, agents need to learn together with other learning agents, creating a non-stationary environment~\citep{laurent_world_2011}.
This poses a challenge to the standard approach of learning from exploration~\citep{sutton_reinforcement_2018}, since agents can easily mistake other agents' exploration as environment randomness~\citep{claus_dynamics_1998}.
A particular challenge presented by the AI Economist is that it presents a two-level learning problem, in which the social planner learns a tax policy simultaneously with agents who learn how to optimize their behavior. In effect, agents face a continuously changing reward function. As a consequence, past optimal behavior might not be optimal at later times, which can present a significant learning challenge.

MARL has been effective in learning emergent cooperation in large-scale experiments on complex environments~\citep{bansal_emergent_2017,jaderberg_human-level_2018,OpenAI_dota}.
Previous MARL algorithms have sought to stabilize multi-agent learning by explicitly modeling missing state or policy information~\citep{lowe_multi-agent_2017,tacchetti_relational_2018,shu_m^3rl:_2018}, or assuming some information is shared between agents, including the internal or global state or rewards~\citep{sunehag_value-decomposition_2017,foerster_learning_2017,peysakhovich_prosocial_2017,hughes_inequity_2018,letcher_stable_2018,balduzzi_mechanics_2018}.

In the present paper, we insist on each agent having a policy that only makes use of information that it can individually observe. To make learning efficient, we allow for weight sharing during training. Learned agent behaviors remain distinct, as a result of distinct local states, for example, location in the world, skill, and endowment of resources. This presents a hybrid approach, improving learning efficiency without assuming information  or state sharing between agents.

Optimal taxation can be seen as a form of {\em reward shaping}, which has found a role in preventing undesired social outcomes in multi-agent systems, such as unsustainable resource collection in tragedy-of-the-commons style social dilemmas \citep{leibo_multi-agent_2017}. Reward shaping has also been shown to induce cooperation in spatiotemporal games \citep{mguni_coordinating_2019,hughes_inequity_2018,jaques_social_2018}.
However, these works do not consider the kinds of economic environments we study here,  do not consider the design of tax policies, and  make use of manually-crafted reward shaping.
\paragraph{Machine learning for Economic Design.}
The problem of {\em automated mechanism design} was first formalized by~\cite{ConitzerS02,ConitzerS04a}, and there are polynomial time algorithms for the design of Bayesian incentive-compatible, optimal auctions~\citep{CaiDW12a, CaiDW12b,CaiDW13}.  
  \cite{dutting_optimal_2019} were the first to study the use of deep machine learning for the design of the allocation and payment rules of revenue-optimal auctions. By insisting on incentive-compatible or approximately incentive-compatible designs, their framework can reproduce known optimal designs and also be applied to problems out of reach of current theory. Subsequent work has also adopted neural networks for the design of optimal auctions in settings with budget-constrained bidders~\citep{FengEtAl18},  for the design of  auctions in settings with payment redistribution~\citep{Tacchetti19}, for single-bidder settings~\citep{STZ18}, as well as for problems of social choice~\citep{GolowichEtAl18}.
The use of machine learning for the design of auction mechanisms was earlier pioneered by~\cite{DuettingFJLLP12}, who studied the design of payment rules for a given allocation rule.
Another line of work explores the sample complexity of the problem of learning an optimal auction, typically focusing on simpler settings~\citep{ColeR14,MorgensternR15,BalcanSV16,GonczarowskiW18}. Earlier work studied  the use of machine learning for the design of voting rules~\citep{procaccia09} and for matching and assignment problems~\citep{Narasimhan_ijcai16,Narasimhan_UAI16}.

In the aforementioned settings, the agents do not learn how to behave. Rather, the economic policies (typically auctions) are designed such that  truthful behavior is optimal for an agent.
This avoids the need for two-level learning, where agent behaviors are learned at the same time as an economic policy is learned. An earlier literature did study the co-evolution of  agent behaviors and economic designs, but without making use of reinforcement learning and without studying  tax polices~\citep{DBLP:conf/sigecom/Byde03,DBLP:conf/amec/PhelpsMPS02,DBLP:journals/aamas/PhelpsMP10}. %
Stackelberg equilibria have also been widely studied in other kinds of sequential environments, especially security games. These are two-level problems where the policy of the first-mover (the defender) induces an environment for the second-mover  (the attacker)~\citep{DBLP:conf/atal/PitaJMOPTWPK08,DBLP:books/daglib/0040483}. Recent work has  adopted MARL for the study of security games~\citep{DBLP:conf/aaai/WangSYWSJF19,DBLP:journals/corr/abs-1911-08799}. Two-stage problems also arise in multi-agent  problems where the behavior of some agents is optimized in order to improve the  overall system behavior~\citep{DBLP:conf/nips/DimitrakakisPRT17,DBLP:conf/nips/CarrollSHGSAD19,tylkin20}.
Other work has made use of reinforcement learning to study resource allocation games such as Blotto~\citep{DBLP:conf/icml/BalduzziGB0PJG19}.

Closest in spirit to the present paper, but used for the design of allocation mechanisms  rather than for tax policy (for example matching sellers to buyer queries at Taobao and for internet advertising at Baidu), is the work of~\cite{DBLP:conf/ijcai/Tang17a} and~\cite{shen20}, who make use of RL to improve market design while also allowing for agent behavior to respond to new rules.
\cite{DBLP:conf/aaai/ThompsonNL17} have also advanced the idea of the ``Positronic Economist" (see also~\cite{DBLP:journals/aamas/VorobeychikRW12} and~\cite{DBLP:conf/sigecom/BunzLS18}), which, borrowing from Asimov's positronic brain, describes a system that can be used to represent and then automatically analyze the equilibria that correspond to the rules of economic mechanisms. \cite{parkes_economic_2015} have written, generally, about the role of economic design in economies in which transactions are increasingly mediated through AI systems.

\subsection{Outline}

In Section~\ref{sec:econsim}, we describe our use of economic simulations and the structure of these simulations. We explain the basic economic drivers and principles that govern the economic AI agents. We then showcase the resulting social outcomes, such as equality and productivity, in such worlds.
In Section~\ref{sec:opttax}, we describe how optimal taxes can shape socioeconomic outcomes, and the central dilemma of balancing equality and productivity. We introduce our RL approach to learning optimal taxes through interaction with economic simulations.
In Section~\ref{sec:expai}, we provide empirical results that validate the effectiveness of the AI Economist in optimizing social outcomes. We analyze the qualitative behavior of AI-driven taxes and economic AI agents.
In Section~\ref{sec:exphuman}, we show that the AI Economist is also effective in experiments on Amazon Mechanical Turk (MTurk), with human participants earning  money.
We conclude in Section \ref{sec:conclusion} with a discussion of future directions, and present our ethical review in Section \ref{sec:ethics}.

\section{Economic Simulations: Learning in Gather-and-Build Games}
\label{sec:econsim}
This section introduces our framework for studying economic design through simulation with AI agents.
We describe the core mechanics of the simulated environment, including the objective that AI agents are trained to optimize, and we describe the emergent behavior that is typical of economic AI agents in this setting.
For ease of exposition, we focus this section on experiments with no taxes applied (``free-market'') in order to illustrate the kinds of social outcomes that taxes may help to correct and some of the challenges faced when designing an optimal taxation scheme.
\subsection{Notation and Preliminaries}
\label{sect:econgames}

In this work, we use notation that borrows from both the reinforcement learning and the optimal tax theory literature, see Table~\ref{tab:notation}.

\begin{table}[h!]
    \begin{minipage}[t]{0.48\linewidth}
        \begin{small}
            \begin{center}
                \begin{tabular}[c]{ll}
                    \multicolumn{1}{c}{\textbf{}} &
                    \multicolumn{1}{c}{\textbf{}} \\
                    $t$ & time \\
                    $i,j,k$ & agent indices \\
                    \hline
                    $\mweight$, $\pweight$ & model weights \\
                    \hline
                    $\st$ & state \\
                    $\ob$ & observation \\
                    $\ac$ & action \\
                    $\rew$ & reward \\
                    $\policy$ & policy \\
                    $\df$ & discount factor \\
                    $\trans$ & state-transition, world dynamics \\
                    \hline
                    $\hidden$ & hidden state
                \end{tabular}
                \label{tab:notation_rl}
            \end{center}
        \end{small}
    \end{minipage}
    \begin{minipage}[t]{0.48\linewidth}
        \begin{small}
            \begin{center}
                \begin{tabular}[c]{ll}
                    \multicolumn{1}{c}{\textbf{}} &
                    \multicolumn{1}{c}{\textbf{}} \\
                    $\endow$ & endowment \\
                    $\money$ & coin \\
                    $\stone$ & stone \\
                    $\wood$ & wood \\
                    \hline
                    $\income$ & income \\
                    $\labor$ & labor \\
                    $\util$ & utility \\
                    \hline
                    $\tax$ & tax \\
                    $\taxrate$ & tax-rate \\
                    \hline
                    $\plannerpolicy$ & planner policy \\
                    $\socialwelfare$  & social welfare \\
                    $\socialwelfareweight$  & social welfare weight \\
                    $\socialmarginalwelfareweight$  & social marginal welfare weight \\
                    \hline
                    $\gini$ & Gini index \\
                    $\equality$ & Equality index \\
                \end{tabular}
                \label{tab:notation_econ}
            \end{center}
        \end{small}
    \end{minipage}
    \caption{Notation. Subscripts are indices. Superscripts are labels.    \label{tab:notation}}
\end{table}

Formally, we build on the framework of partial-observable multi-agent Markov Games (MGs) \citep{sutton2018reinforcement}, defined by the tuple $(S, A, \rew, \trans, \df, \ob, \calI)$, where $S$ and $A$ are the state and action spaces, respectively, and $\calI$ are agent indices.
Bold-faced quantities denote vectors, e.g., $\bm{a}=(a_1,\ldots,a_N)$, the action profile for $N$ agents.
MGs proceed in episodes that last $\eplen + 1$ steps (possibly infinite), covering $\eplen$ transitions.
At each time $t \in [0,\eplen]$, the world state is denoted $\st_t$. Each agent $i = 1,\ldots,N$ receives an observation $\ob_{i,t}$, executes an action $\ac_{i,t}$ and receives a reward $\rew_{i,t}$.
The environment transitions to the next state $\st_{t+1}$, according to the transition distribution $\trans\brck{\st_{t+1} | \st_t, \va_t }$.
Agent-specific observations $\ob_{i, t}$ describe the portion of the state $\st_t$ that agent $i$ is able to observe.

Each agent learns a policy $\pol_i\brck{\ac_{i,t} | \ob_{i,t}, \hidden_{i,t}; \theta_i}$ that maximizes its $\df$-discounted expected return, where the policy is conditioned on the history of past observations by maintaining a hidden state $\hidden_{i, t}$, and  where
$\theta_i$ parameterizes the policy of agent $i$.
Let $\bm{\pol} = \brck{\pol_1,\ldots,\pol_N}$ denote the joint policy
and $\bm{\pol}_{-i} = \brck{\pol_1,\ldots,\pol_{i-1},\pol_{i+1},\ldots,\pol_N}$ denote the  policy without agent $i$.
Through reinforcement learning, agent $i$ seeks a policy to solve
\eq{\label{eq:rl-agent-objective}
\max_{\theta_i} \E_{a_i\sim \pi_i, \va_{-i}\sim\bm{\pi}_{-i}, \st'\sim\trans} \brcksq{ \sum_t \df^t \rew_{i,t}},
}
for discount factor $\gamma \in (0, 1)$. Equation \ref{eq:rl-agent-objective} describes an agent $i$ that maximizes its expected reward, which depends on the behavioral policies $\bm{\pi}_{-i}$ of the other agents and the environment transition dynamics $\trans$.
This is a policy that best responds to the policies of other agents, given the dynamics of the simulated environment and an agent's observations.

For data efficiency, all agents share the same parameters during training,  denoted  $\mweight$, but condition their policy $\policy_i\brck{\ac_i|\ob_i, \hidden_i; \mweight}$ on agent-specific observations $\ob_i$ and  hidden-state $\hidden_i$.
In effect, if one agent learns a useful new behavior for some part of the state space then this becomes available to another agent.
At the same time, agent behaviors remain heterogeneous because they have different observations and hidden states.

\begin{figure}
    \begin{small}
        \begin{center}
            \includegraphics[width=0.4\textwidth]{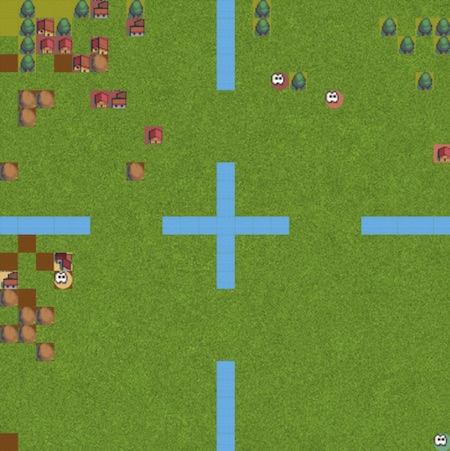}
        \end{center}
        \caption{The Gather-and-Build game. Agents move around, collect resources (wood and stone) and build houses. Agents cannot move through each others' houses, or move through water. Agents can  trade resources.}
        \label{fig:simulation_human_gui}
    \end{small}
\end{figure}

\subsection{Environment Rules and Dynamics}
\label{sec:env_rules_and_dynamics}
\paragraph{Overview.}
We introduce the \textit{Gather-and-Build} game, a two-dimensional grid-world in which agents can move to collect resources, earn coins by using the resources of stone and wood to build houses, and trade with other agents to exchange resources for coins.
Stone and wood stochastically spawn on special resource regeneration tiles.
Agents can move around the environment to gather these resources from populated resource tiles that  remain empty after harvesting until some new resources spawn.

Agents can choose to use one unit of wood and one unit of stone to construct a house, and this places a house tile at the agent's current location and earns the agent some number of coins.
The number of coins earned per house depends on the {\em skill} of an agent, and skill is different across agents. In addition, agents start at different initial locations in the world.
These heterogeneities are the main driver of both economic inequality and specialization in our environment.

Agents can also trade resources, by submitting the number of coins they are willing to accept (an {\em ask}) or are willing to pay (a {\em bid}), respectively, to an open market, for each of wood and stone.
We provide a detailed description of the environment and its underlying dynamics in the appendix (Section~\ref{sec:details_of_environment}).

\paragraph{Labor and Skill.}
Over the course of an {\em episode} (a single play out of the environment), agents  accumulate labor cost, which reflects the amount of effort associated with the actions taken by the agent.
Each type of action (moving, gathering, trading, and building) is associated with a specific labor cost.
Each time an agent performs one of these actions, its accumulated labor is incremented by the action's associated labor cost.
As described below (Section \ref{sec:agent_utility}), agent rewards depend positively on accumulated coin and negatively on accumulated labor.
The labor costs associated with each action type are calibrated so that agents need to be strategic in how they choose to earn income, and all agents experience the same labor costs.

Following taxation theory, we allow agents in the environment to vary by  skill, which describes how much income an agent is able to earn per unit of labor.
We capture this by providing, separately for each agent, (1) a multiplier on the default number of coins earned from building a house, and (2) the probability of gaining bonus resources when harvesting.
The coin payoff for a house depends linearly on skill.
An agent receives a minimum of 10 coin per house built.

A {\em building skill} of 1 (the minimum value) means the agent earns this minimum payoff and a building skill of 2.5 means the agent receives 25 coin per house. The maximum skill value is 3.
An agent's {\em collection skill} is equal to the average number of resources it receives each time it steps on a populated stone or wood resource tile.
As an example, for an agent with a collection skill of 1.2, it will always receive at least 1 resource from stepping on a populated tile and there is a 20\% chance it will also receive a bonus unit of the collected resource.
The minimum collection skill is 1 and the maximum is 2, ranging from never receiving bonus resource units to always receiving them.

We conceptualize the coins that are generated when building a house as coming from a part of the wider economy that our simulation does not directly model. An agent's building skill--- the coin the agent receives from building ---reflects the value that this external market places on a particular agent's houses. The total quantity of coins generated by the simulated agents during an episode reflects the value created by their collective labor.

\paragraph{Environment Scenario.}
All experiments were carried out using the specific world map shown in Figure~\ref{fig:simulation_human_gui}, which has four quadrants,  mostly separated by water  from each other (this blocks movement), and with spatially clustered resources.
We focus on games with four agents, and apply a fixed set of building skills,
 chosen as the means of the quartiles of a clipped Pareto distribution with exponent $a=4$ and scale $m=1$.
Skills and starting locations are randomly assigned to agents.
These building skills correspond to payoffs of 11.3, 13.3, 16.5, and 22.2 coins per house.
In all  experiments, we used episodes of length $\eplen = 1000$ time steps.

\begin{figure}[t!]
\begin{small}
    \begin{center}
    \includegraphics[width=0.617\linewidth]{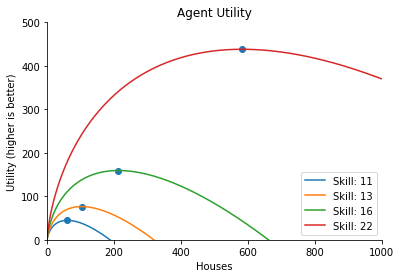}
    \end{center}
    \caption{Agent utility, as defined in Equation \ref{eq:agent_utility}, in a simplified setting where utility only depends on the number of houses built, for four agents with different skills $\skill_0 < \ldots < \skill_3$. For clarity of this visualization, each agent is assumed to receive a fixed income per house, which increases with skill, as described in Section \ref{sec:env_rules_and_dynamics}. For each house built, each agent is also assumed to have exerted a fixed amount of labor. Hence, in this simple example, agent utility only depends on the number of houses built. Each agent experiences a law of diminishing returns (marginal utility decreases as income grows). As agent skill increases, the point of maximal utility is reached at a higher number of houses built.}
    \label{fig:agent_utility}
\end{small}
\end{figure}

\subsection{Using Machine Learning to Optimize Agent Behavior}
\label{sec:agent_utility}

To ground our simulation in economic theory, we model the reward that the agents learn to optimize as a {\em utility function}.
Recall that $\endow_{i,t}$ denotes the endowment of resources (stone and wood) and coin owned by an agent at time $t$.  At time $t$, the utility experienced by an agent $i$ is a function of the number of coins that it has accumulated $\money_{i,t}$, and the total labor that it has exerted $\labor_{i,t}$.
In particular, we adopt a  utility function that is concave and increasing in money, and linearly decreasing in labor:
\eq{\label{eq:agent_utility}
    \util_{i}(\endow_{i,t}, \labor_{i,t}) = \crra\brck{\mon{i}{t}} - \labor_{i, t},
    \quad
    \crra\brck{\income} = \fr{\income^{1-\isoeta}-1}{1-\isoeta}, \quad
    \isoeta > 0.
}

Here,
$\labor_{i, t}$ is the cumulative labor associated with the actions taken by the agent up to time $t$,
and the concave isoelastic utility $\crra$ models diminishing marginal utility over money \citep{debreu_representation_1968}. Parameter $\isoeta$ controls the degree of nonlinearity: higher  $\isoeta$ represents larger deviations from linear behavior.
We assume that all agents share the same form of utility function. This utility function is visualized in Figure \ref{fig:agent_utility} for a simple setting without trading and where there is no labor cost associated with moving or gathering resources.

Rational economic agents optimize their total discounted utility over time, with
\eq{\label{eq:agentproblem}
    \forall i:
    \max_{\policy_i}
    \E_{
    \ac_i    \sim \policy_i,
    \Ac_{-i} \sim \bm{\policy}_{-i},
    \st'     \sim \trans
    }\brcksq{
        \sum_{t = 1}^\eplen \df^t
        \underbrace{\brck{
            \util_i( \endow_{i,t}, \labor_{i,t} )
            - \util_i( \endow_{i,t-1}, \labor_{i,t-1} )
        }}_{=\hspace{2pt} \rew_{i,t}}
        + \util_i( \endow_{i,0}, \labor_{i,0} )
    }.
}

In the paradigm of RL, this is achieved  by defining the instantaneous reward $\rew_{i,t}$ for agent $i$ as the change in utility of  agent $i$   at time $t$.

Equation \ref{eq:agentproblem} describes a multi-agent optimization problem, when the agents are simultaneously optimizing their behavior, since the utility for agent $i$ depends on the behaviors of other agents (for example, their gathering, building and trading actions).
For instance, another agent might block an agent's access to resources, which would impact how many houses the agent can build in the future and hence its future utility.

In general, such optimization problems are described as partially-observable multi-agent Markov games, and optimal solutions correspond to refinements of Nash equilibria.  A set of policies form a Nash equilibrium as long as no agent wants to unilaterally deviate from its own policy. Refinements such as  subgame-perfect equilibria also require rational, off-equilibrium behavior. Although computing equilibria for complex environments such as this remains out-of-reach,  we will see that RL can nevertheless be used to achieve sensible, emergent behaviors (and behaviors that also drive good tax policy, when coupled with the use of the AI Economist).

\paragraph{Deep RL agents.}
We make use of a deep neural network to model agent policies,
\eq{
 \ac_{i, t} \sim \policy(\ob^\text{world}_{i, t}, \ob^\text{agent}_{i, t}, \ob^\text{market}_{i, t}, \ob^\text{tax}_{i, t}, \hidden_{i, t-1}; \mweight).
}

The output of this policy network includes a probability distribution over actions, with $\ac_{i, t}$ sampled from this distribution. Not represented in the notation, the policy network also generates an updated hidden state $\hidden_{i, t}$.
The inputs to the network include the agent-specific hidden state and agent-specific observations, which are decomposed as follows:
\begin{itemize}
    \item $\ob^\text{world}_{i, t}$: spatial observations from the nearby world.
    \item $\ob^\text{agent}_{i, t}$: the \textit{public} agent state (such as resource and coin endowments), as well as the \textit{private} agent state (such as skill values and labor performed).
    \item $\ob^\text{market}_{i, t}$: the full market state, including available offers to buy/sell resources.
    \item $\ob^\text{tax}_{i, t}$: the tax rates in effect.\footnote{We include tax information even for the free-market, when all tax rates are zero. This ensures that the structure of the observations is the same for all taxation schemes.}
\end{itemize}

\begin{figure}[t!]
    \begin{small}
        \begin{center}
            \includegraphics[width=0.8\textwidth]{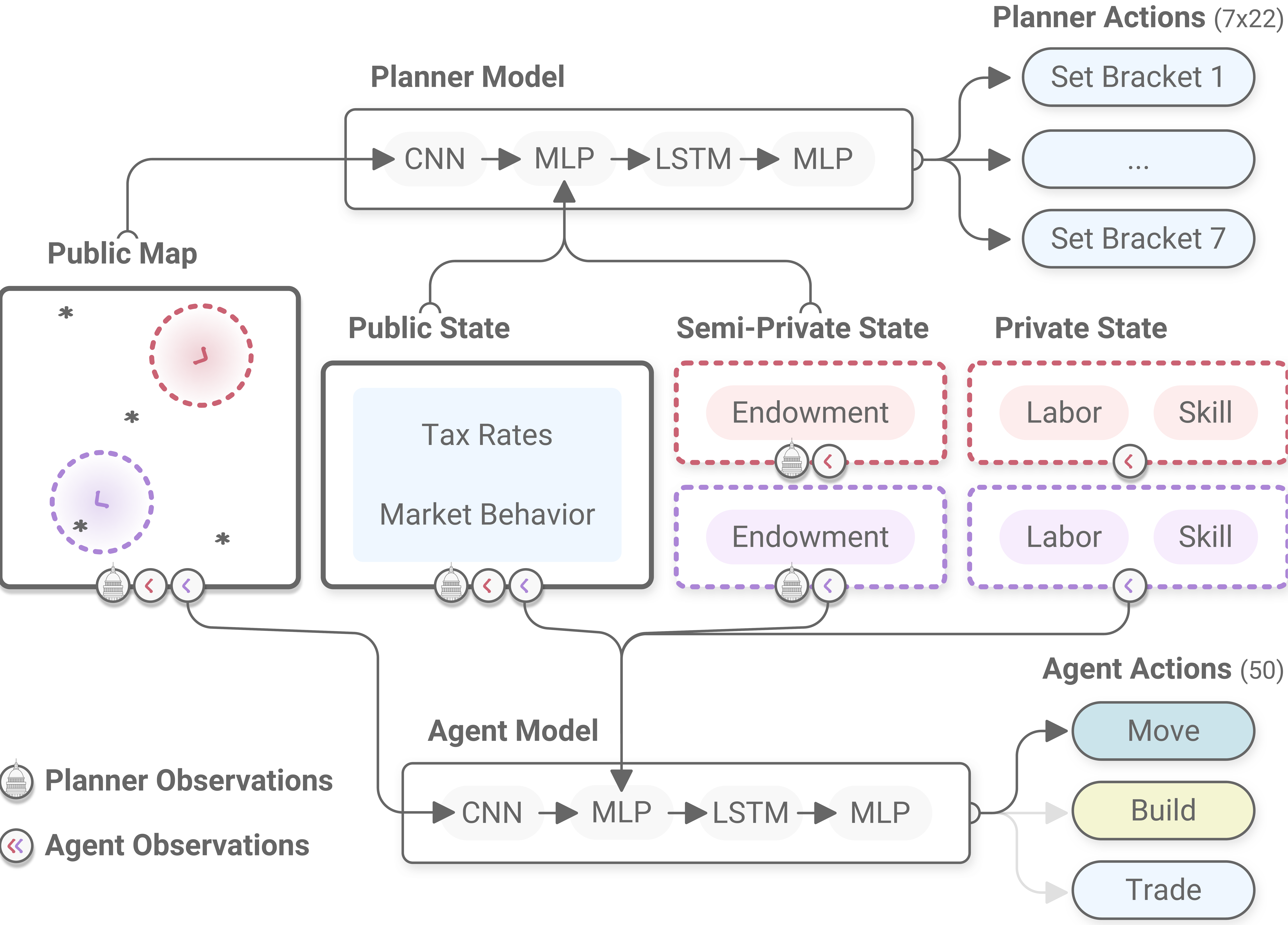}
        \end{center}
        \caption{
        Schematic overview of the general network architecture used in our work.
        Spatial observations are processed by a stack of two convolutional layers (CNN) and flattened into a fixed-length feature vector.
        This feature vector is concatenated with the remaining observation inputs and the result is processed by a stack of two fully connected layers (MLP).
        The output is then used to update the hidden state of an LSTM and action logits are computed via a linear projection of the updated hidden state.
        Finally, the network computes a softmax probability layer for each action head.
        For the agent policy, there is a single action space and action head.
        For the tax policy, there is a separate action space and action head for each tax rate the tax policy controls (described below).
        }
        \label{fig:neural_network_policies}
    \end{small}
\end{figure}

Section \ref{sec:details_of_environment} of the appendix provides exact details of the information available in the agents' observations.
The learned hidden state $\hidden_{i,t}$ is used to encode the history of past observations.
Figure \ref{fig:neural_network_policies} depicts the general network architecture used here.

\begin{figure}[t!]
    \begin{center}
        \includegraphics[width=0.95\textwidth]{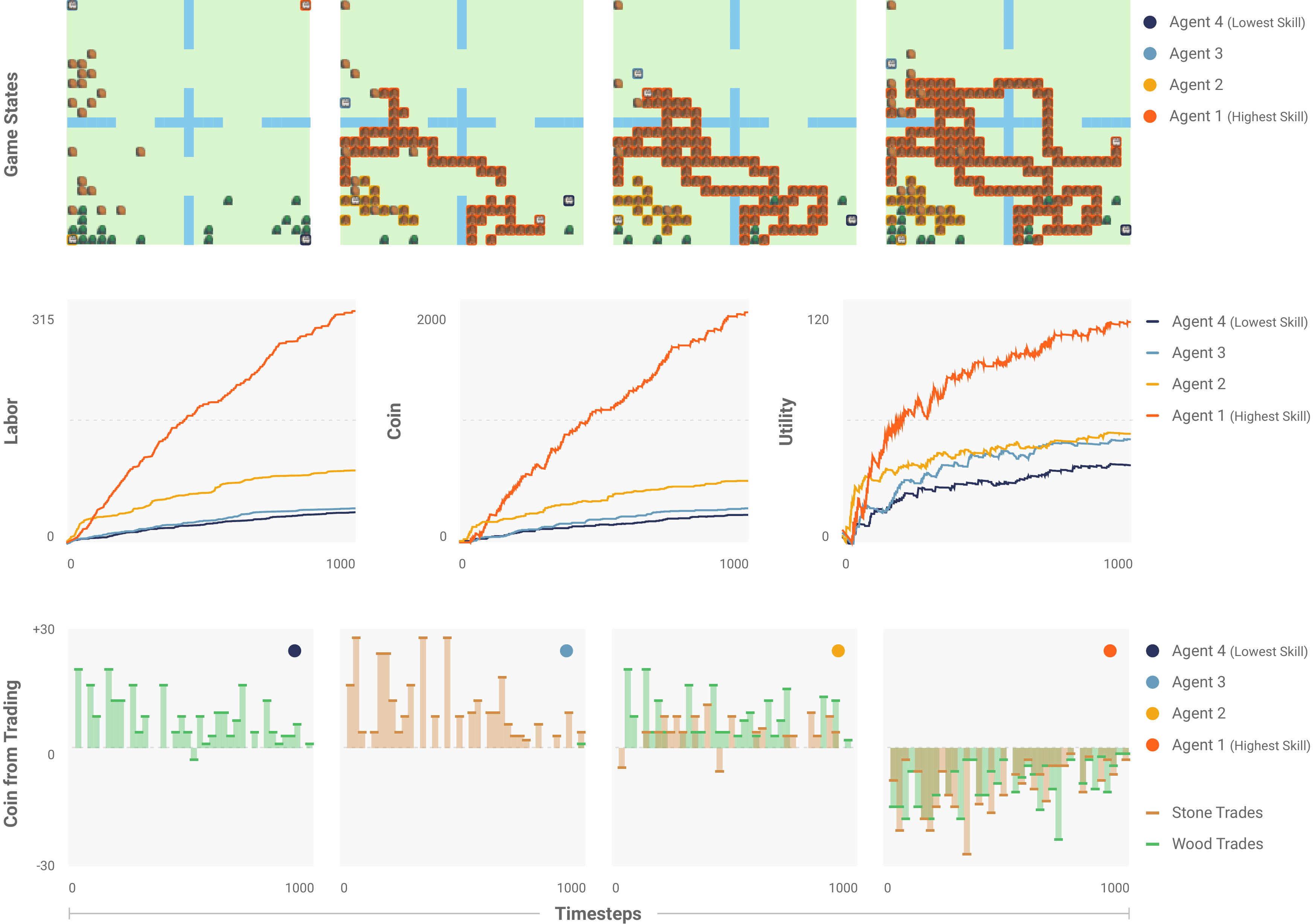}
  \end{center}
  \caption{
  Breakdown of an example rollout in a no-tax environment.
  Agents labels are sorted according to each agent's building skill (dark-blue is lowest, orange is highest). These correspond to agent payoffs of 11.3, 13.3, 16.5, and 22.2 coins per house.
  The top panels illustrate the world state, including agent locations, available resources, and built houses, as the episode progresses.
  The middle panels illustrate the accumulated labor, coin, and utility of each agent over the episode.
  Highly skilled agents ultimately experience more utility.
  The bottom panels illustrate the net coin received/spent from trading over the episode, for each of the four agents.
  Each bar represents the net coin within a window of 25 timesteps, with upward bars indicating net income (agent predominantly sold), downward bars indicating net cost (agent predominantly bought), and color indicating the resource type.
  Agents with lower building skill choose to earn income through gathering resources and selling them to the highly skilled agents.
  }
  \label{fig:rollout}
\end{figure}

\paragraph{Emergent Behavior of AI Agents.}
Figure \ref{fig:rollout} provides a breakdown of an example rollout of play by AI agents across a single episode, once training has proceeded for a large number of episodes.
Each agent has a unique color. Agents are ordered from low to high skill as dark-blue, light-blue, yellow, and orange, corresponding to payoffs of 11.3, 13.3, 16.5, and 22.2 coins per house, respectively.

This rollout reveals an interesting specialization of effort.
The dark- and light-blue agents focus entirely on collecting wood and stone (respectively), the orange agent focuses almost entirely on building houses, and the yellow agent builds several houses early on before switching to collecting and selling.

This pattern of behavior and division of labor is typical of agents trained in this simulated environment, and stems from the different incomes each agent can earn per house it builds, as well as the agents' initial locations in the world.
In particular, the low skilled dark- and light-blue agents learn to shift their strategies entirely away from building houses.
These agents earn their income by selling resources to the higher skilled agents, who choose to earn income through building (Figure \ref{fig:rollout}, middle).
The yellow agent earns enough income from building to do so early on, making use of the nearby resources, but then switches strategies.

This specialization is a consequence of agents learning to maximize their own individual objectives.
We do not impose these roles or behaviors directly. Rather, this  specialization arises as a result of differently skilled workers learning to balance their income and effort.
This emergent behavior helps to validate the framework as an economic simulation, by reproducing a standard feature of real world economies, that of specialization.
Standard economic intuition states that agents should specialize in whichever means of production allows them to most efficiently convert their labor to income, and this is consistent with the behaviors that the AI agents discover.

Even with specialization, the agents' incomes can vary considerably.
While a free-market economy maximizes productivity, it provides no guarantee on income equality.
This is evident in the highly unequal incomes  experienced by the AI agents.

\section{Machine Learning for Optimal Tax Policies}
\label{sec:opttax}
We now introduce a \emph{social planner} who uses economic policy to improve social outcomes, in particular taxation together with  redistribution. The challenge is that taxation can reduce productivity. Workers may choose to forgo labor as a result of paying tax on income, and thus gaining less utility for labor effort. This may have a particularly strong effect on the higher skilled and thus more productive workers. Thus, there is a trade-off between equality and productivity: the same interventions that allow wealth to be redistributed also result in there being less wealth to redistribute in the first place.
As a result of this coupling between taxation and labor, determining an optimal tax policy poses a difficult, constrained optimization problem.

\begin{figure}
 \begin{small}
 \begin{center}
  \includegraphics[width=\textwidth]{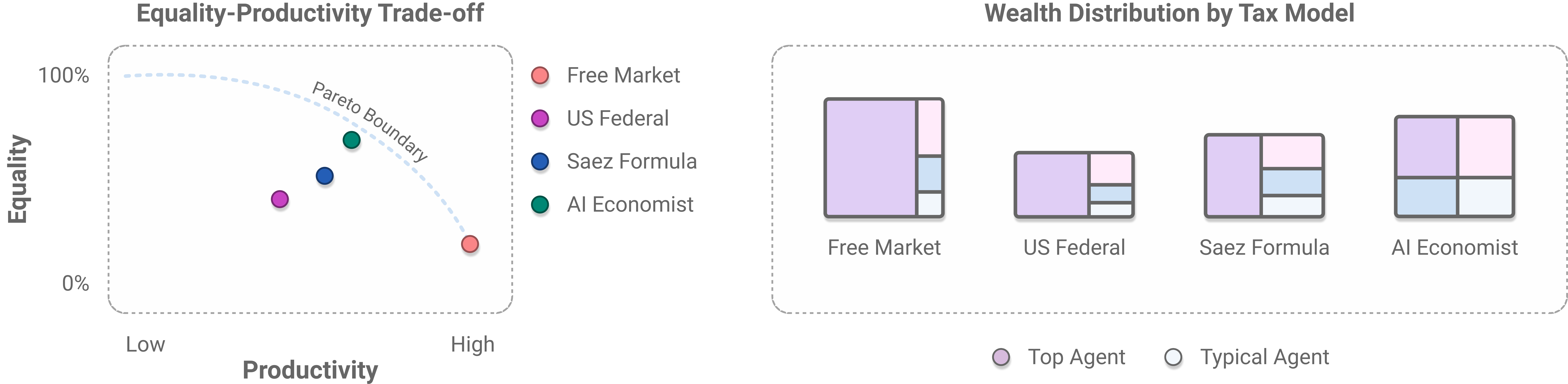}
 \end{center}
 \caption{A conceptual view of how taxes can impact social outcomes. Left: Taxes can improve equality by transferring wealth. However, taxes can also decrease productivity, because they can discourage work. The AI Economist seeks a tax policy that optimizes this trade-off. The Pareto boundary is the set of maximal trade-offs. Right: Taxes impact productivity (total income, represented by the area of the big squares), and equality (the relative difference in sizes of the smaller squares). The AI Economist achieves the best trade-off (measured in equality times productivity).
 }
 \label{fig:eq_prod_tradeoff}
\end{small}
\end{figure}

A conceptual view of the trade-off between productivity and equality for different tax policies is illustrated in Figure~\ref{fig:eq_prod_tradeoff}.
The spectrum of tax policies has two extremes: the free market, which only considers productivity and does not raise any taxes; and pure redistribution, which divides all incomes equally amongst all workers and thus achieves equality but at the potential cost of a large drop in productivity.
Here, the notion of optimality implies that a tax policy realizes a trade-off between equality and productivity along the Pareto boundary linking these two extremes.
The optimal tax literature has proposed several solutions, including the tax formula proposed by \cite{saez_using_2001} (shown here, together with the AI Economist, with  purely illustrative tradeoffs).
But the results from optimal tax policy are limited to simple economic models, and require various simplifying assumptions, for example about the effect of higher taxes on labor choices.

The remainder of this section describes our approach for studying optimal taxation.
We describe the kind of tax policy learned by the AI Economist, define the types of social objectives that can be adopted, and describe how we use reinforcement learning to jointly optimize agent behavior as well as the tax policy used in the economy.

\subsection{Periodic Taxes with Bracketed Schedules}
\label{sect:brackettax}

\paragraph{Income Taxes.}
There are many possible choices for tax policies.
In this work, we focus on periodic income taxes with lump-sum redistribution.
Each {\em tax period} lasts $\taxperiodlen$ steps (we use $\taxperiodlen=H/10$, so that there are ten tax periods per episode).
The taxes in period $p$, beginning at time step $t$ and ending at time step $t+\taxperiodlen$, are applied to the income $\income^p_i$ earned by an agent $i$ within that tax period.

At the start of each tax period, the planner chooses a {\em tax schedule} $\incometax(\income)$ that specifies the amount of taxes an agent will owe as a function of the income it earns during the period.
At the end of each tax period the total tax revenue is evenly redistributed back to the agents, so that the adjusted, post-tax income to agent $i$ in period $p$ is given by
\eq{
 \posttax{\income}^p_i = \income^p_i - \incometax(\income^p_i) + \fr{1}{N} \sum_{j=1}^N \incometax(\income^p_j).
 \label{eq:posttaxincome}
}

At the end of an episode, each agent's coin endowment is the sum of its post-tax incomes in each period:
$\money_{i,H} = \ \sum_p \posttax{\income}^p_i$.

\paragraph{Bracketed Tax Schedules.}
To allow comparison across different schemes, we adopt income brackets for describing a tax schedule, imitating the US federal taxation scheme.
A bracketed schedule defines a set of {\em cut-off income levels} $\bracketcutoff_b$, where $b = 0, \ldots B$, for $B$ income brackets.
The edges of bracket $b$ are $[\bracketcutoff_b, \bracketcutoff_{b+1}]$, and, by definition, $\bracketcutoff_0 = 0$ and $\bracketcutoff_B = \infty$.
The social planner sets the tax schedule $\incometax(\income)$ by choosing the \textit{marginal tax rate} $\mtaxrate \in [0, 1]^B$ to be applied within each bracket.

Given this, the total tax payment $\incometax(\income)$ for an agent earning $z$ in a tax period is determined by taking the sum of the amount of income within each bracket $[\bracketcutoff_b, \bracketcutoff_{b+1}]$ times that bracket's marginal rate $\mtaxrate_b$:
\eq{
 \incometax(\income) = \sum_{b=0}^{B-1} \mtaxrate_b \cdot \brck{
 \brck{ \bracketcutoff_{b+1} - \bracketcutoff_b } \bm{1}[ \income > \bracketcutoff_{b+1} ]
 + \brck{ \income - \bracketcutoff_b } \bm{1}[ \bracketcutoff_b < \income \leq \bracketcutoff_{b+1} ]
 },
}
where $\bm{1}[ \income > \bracketcutoff_{b+1} ]\in \{0,1\}$ is an indicator function for whether $\income$ saturates bracket $b$ and $\bm{1}[ \bracketcutoff_b < \income \leq \bracketcutoff_{b+1} ]\in \{0,1\}$ is an indicator function for whether $\income$ falls within bracket $b$.
\subsection{Optimal Taxation}
\label{sect:socialwelfarefunctions}
\paragraph{Social Welfare Functions.}
The objective of optimal tax theory is described through a \emph{social welfare function} $\socialwelfare{}$.
Social welfare can be expressed in many ways. One approach considers the trade-off between income equality and productivity.
For this, the {\em equality} in an economy at some point in time can be defined as the complement of the {\em normalized Gini index } on the distribution on wealth, this wealth defined as the cumulative number of coins owned by an agent after taxation and redistribution.

For an agent population with monetary endowments $\Money=(\money_1,\ldots,\money_{N})$, we define equality $\equality(\Money)$ as:
\begin{align}
 \equality(\Money) = 1 - \gini(\Money) \fr{N}{N-1},& \quad 0 \leq \equality(\Money) \leq 1,
\end{align}
where the Gini index is defined as,
\begin{align}
\gini(\Money) = \fr{\sum_{i=1}^N \sum_{j=1}^N |\money_i - \money_j|}{ 2N \sum_{i=1}^N \money_i},& \quad 0 \leq \gini(\Money) \leq \fr{N-1}{N}.
\end{align}

Given this, $\equality = 1$ implies perfect equality (all endowments of money are identical), while $\equality = 0$ means perfect inequality (one agent owns all money).
The {\em productivity} in an economy at some point in time is defined as the sum of all wealth over all agents:
\begin{align}
 \productivity(\Money) = \sum_{i=1}^N \money_i.
\end{align}
We write $\equality_t(\Money_t)$ and $\productivity_t(\Money_t)$ to denote the equality and productivity, respectively, based on the cumulative endowment $\Money_t$ up to  time $t$.

The primary social welfare function that we consider in this work  optimizes a trade-off between equality and productivity, defined as the product of equality and productivity:
\begin{align}
 \label{eq:eqprodswf}
 \socialwelfare_t(\Money_t) = \equality_t(\Money_t) \cdot \productivity_t(\Money_t).
\end{align}

Another family of social welfare functions, and one that receives attention in the  optimal taxation theory, is the family of linear-weighted sums of agent utilities, defined for  weights $\socialwelfareweight_i\geq 0$:
\eq{\label{eq:inv_income_weighted_utility}
\socialwelfare_t(\Money_t,\vlabor_t) = \sum_{i=1}^N \socialwelfareweight_i \cdot \util_i\brck{ \money_{i,t}, \labor_{i,t} }.
}

 Some illustrative choices for the weights adopted in this social welfare function include:
\begin{itemize}
 \item Utilitarian: $\socialwelfareweight_i=1$, indicating no preference for any agent
 \item Rawlsian: $\socialwelfareweight_i=\bm{1}[\money_{i,t} = \min_{j\in\mathcal{J}}\money_{j,t}]$, which  focuses on the poorest agents
 \item Inverse income-weighted: $\socialwelfareweight_i=1/\money_{i,t}$, which preferences the agents with lower endowments over those with higher endowments.
\end{itemize}

In this work, we will  mainly make use of the product of equality and productivity as the social welfare function, and it is this that the AI Economist is configured to optimize for. But many other choices are possible.
A key benefit of our framework is that it is compatible with any social welfare function.

For the purposes of comparing the performance of the AI Economist and other tax frameworks, we also adopt a variation on the second family of social welfare functions, where we adopt inverse income-weighted  weights and consider agents' cumulative endowment at the end of an episode.

\paragraph{The Planner's Problem.}
The planner can observe agent endowments, $\vendow_{i,t}$, as well as the global state of the world, including agent positions, available resources, and market states. The planner cannot directly observe agent skills or other endogenous values such as labor or utility functions. The planner does not personalize taxes, but adopts a single tax schedule for all agents.

Similar to the agent policies, the planner can make use of the entire history of observations, via a learned hidden state, to implement its tax policy.
Based on the information available, it adopts a {\em tax policy} $\plannerpolicy$ to set tax rates $\mtaxrate$ in any given tax period.
The planner's objective is to optimize social welfare,
\eq{\label{eq:plannerproblem}
 \max_{\plannerpolicy}
 \E_{
 \mtaxrate \sim \plannerpolicy,
 \Ac \sim \bm{\policy},
 \st' \sim \trans
 }
 \brcksq{
 \sum_{t=1}^H \df^t \underbrace{
 \brck{\socialwelfare_t - \socialwelfare_{t-1}}}_{=\hspace{1pt}\rew_{p,t}}
 + \socialwelfare_0
 },
}
where $\socialwelfare_t$ is used to denote the social welfare at time $t$, based on cumulative endowment and labor up until that time, and the planner's instantaneous reward $\rew_{p,t}$ is the change in social welfare at time $t$.
Without discounting, this objective reduces to the total social welfare at the end of an episode.

This planning problem includes the effect of agents' behavior, encoded through agent policies $\bm{\pol}$, and this behavior of agents depends on the tax policy.
As such,~\eqref{eq:plannerproblem} encodes a difficult optimization problem; because taxes affect agents' income and thus utilities, the planner is effectively changing the agents' MDP.\footnote{The problem has the flavor of a Stackelberg game in which  the planner is a first-mover (Stackelberg leader), announcing a tax schedule, and where the agents responding to this tax schedule (Stackelberg followers).}

\begin{figure}
    \begin{small}
        \begin{center}
            \includegraphics[width=0.95\textwidth]{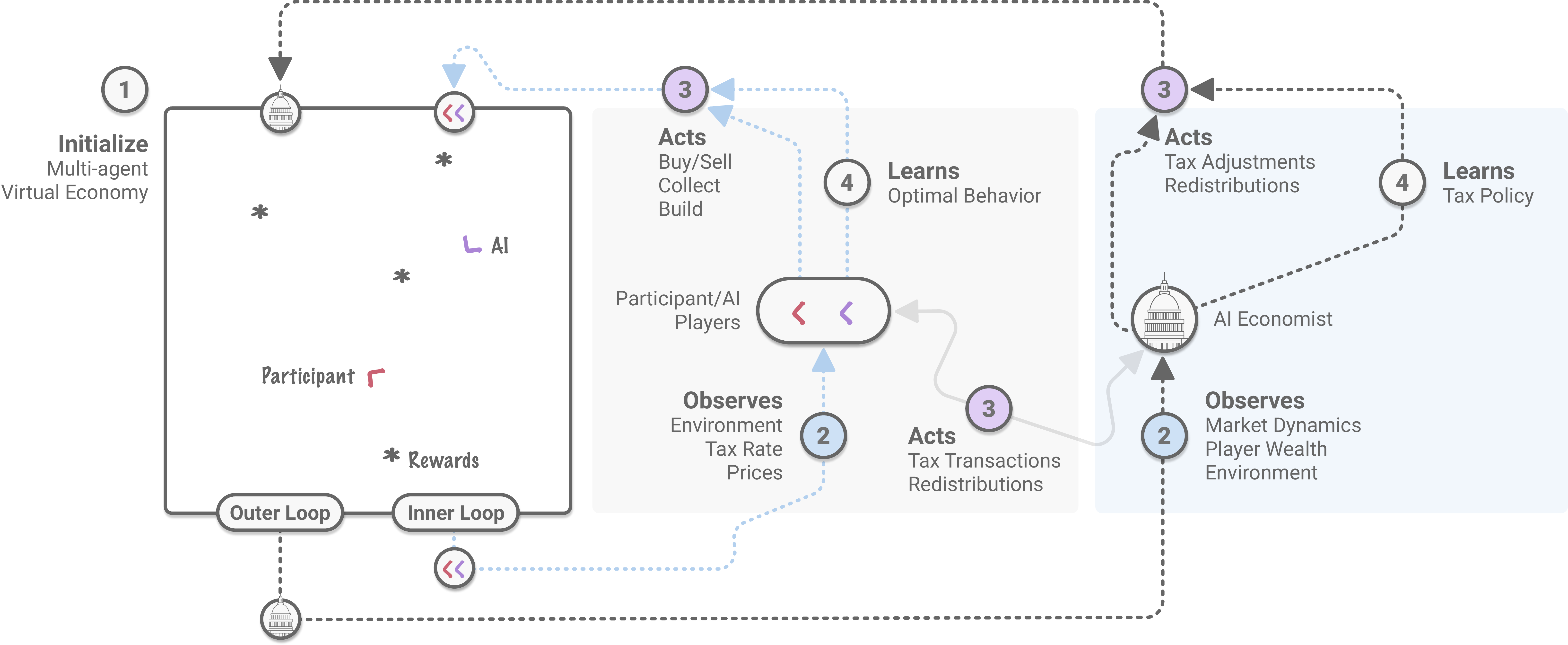}
        \end{center}
        \caption{Two-level RL. In the inner loop, RL agents gain experience by performing labor, receiving income, and paying taxes, and learn through balancing exploration and exploitation how to adapt their behavior to maximize their utility. In the outer loop, the social planner adapts tax policies to optimize its social objective.}
        \label{fig:innerouterrl}
    \end{small}
\end{figure}
\subsection{Inner-Outer-Loop Reinforcement Learning}
\label{sect:innerouterrl}
Reinforcement learning provides a suitable learning framework for adapting behaviors in a sequential environment, and is used here to allow both the economic agents and the social planner to learn from experience collected through trial-and-error strategies. 
We conceptualize the RL framework used in this paper as including two levels of learning, namely an `inner' loop and an `outer' loop, as depicted in Figure~\ref{fig:innerouterrl}.

In short, the key benefits of our RL framework are:
\begin{itemize}
    \item the social planner can optimize taxes for any social objective \socialwelfare, and
    \item given a choice of social welfare functions $\socialwelfare$, the social planner does not need prior world knowledge, prior economic knowledge, or assumptions on the behavior of economic agents.
\end{itemize}

The main challenge posed by this two-level RL problem comes from the fact that each level effectively determines the MDP faced by the other level.
As the planner learns and changes taxes, the agents' utility and reward landscapes change.
In turn, as agents learn and adapt to new taxes, their behavior changes the expected social welfare generated through the tax schedule.
In this way, simultaneous learning creates an unstable reward landscape for both the  agents and the planner.
\paragraph{Inner Loop.}
In the inner loop, RL agents gain experience by performing labor, receiving income, and paying taxes, and learn through trial-and-error how to adapt their behavior to maximize their utility. Given a fixed tax policy, this is a standard RL problem in which agents iteratively explore and discover which behaviors are optimal for their fixed utility function, while observing the active tax schedule.

However, because the tax policy is changing, and in turn the behavior of others, the agent is faced with a non-stationary MDP.
Specifically, the utility of agent $i$ depends on its post-tax incomes $\endow_i$ (Eq \ref{eq:posttaxincome}).
The non-stationarity faced by agents can be understood by considering their learning objective in the context of a changing tax policy (generalizing Eq \ref{eq:agentproblem}):
\eq{\label{eq:agentproblemposttax}
    \max_{\policy_i}
    \E_{
    \ac_i    \sim \policy_i,
    \Ac_{-i} \sim \V{\policy}_{-i},
    \mtaxrate \sim \plannerpolicy,
    \st'\sim\trans
    }\brcksq{
        \sum_{t = 1}^H \df^t \brck{
            \util_i( \endow_{i,t}, \labor_{i,t} )
            - \util_i( \endow_{i,t-1}, \labor_{i,t-1} )
        } + \util_i( \endow_{i,0}, \labor_{i,0} )
    }.
}

The agent's expected future utility  is conditional on both the current state and the current tax schedule.
Hence, as the planner's policy $\plannerpolicy$ changes, the taxes  that an agent experiences will change, and agents face a non-stationary learning environment in which they constantly need to adapt to a changing utility landscape.
As time goes on, because the post-tax income for the same type and amount of labor can change over time, agent decisions that were optimal in the past might not be optimal in the present.

\paragraph{Outer Loop.}
In the outer loop, the social planner adapts its tax policy to optimize the social objective, following the learning objective defined in Eq \ref{eq:plannerproblem}.
Since the agents also change their behavior, the planner also faces a non-stationary problem, due to the dependence of Eq~\ref{eq:plannerproblem} on the policy of each agent.

In order to allow both  the agents and the planner to learn an optimal behavior, which considers the best response of agents to the tax policy, the agents and the planner must be trained jointly.
That is, there is little point to training a planner using a set of fixed agent policies, since the social welfare achieved  would not be meaningful without considering the way agents' behaviors would change in response.

It should also be pointed out that our terminology is not meant to imply any nested training structure.
When learning a tax policy, we train the agent policies and planner policies jointly, following standard practice for multi-agent RL. Here, joint training entails both agents and the planner updating their weights simultaneously during each training episode.
Algorithm \ref{alg:innerouterrl} in the Appendix provides a more detailed description of the training framework.
\begin{algorithm}[t]
    \floatname{algorithm}{Algorithm}
    \caption{Inner-Outer Loop Reinforcement Learning. Economic agents and social planner learn simultaneously. Bold-faced symbols indicate quantities for multiple agents. Note that agents share weights.}
    \label{alg:innerouterrl}
\begin{algorithmic}
    \Require Sampling horizon $\samplinghorizon$, tax period length $\taxperiodlen$
    \Require On-policy learning algorithm $\sA$ (for instance, A3C, PPO)
    \Require Stopping criterion $C$ (for instance, agent and planner rewards have not improved)
    \Ensure Trained agent and planner policy weights $\theta, \phi$
    \State $\st, \V{\ob}, \ob_p, \V{\hidden}, \hidden_p \Leftarrow \st_0, \V{\ob}_0, \ob_{p, 0}, \V{\hidden}_0, \hidden_{p, 0}$ \Comment{Reset episode}
    \State $\theta, \phi \Leftarrow \theta_0, \phi_0$ \Comment{Initial agent and planner policy weights}
    \State $D, D_p \Leftarrow \{\}, \{\}$ \Comment{Reset agent and planner transition buffers}
    \While{training}
    \For{$t = 1, \ldots, \samplinghorizon$}
        \State $\Ac, \V{\hidden} \Leftarrow \V{\policy}(\cdot | \V{\ob}, \V{\hidden}, \mweight)$ \Comment{Sample agent actions; update hidden state}
        \If{$t$ mod $\taxperiodlen$ = 0}  \Comment{First timestep of tax period}
            \State $\mtaxrate, \hidden_p \Leftarrow \plannerpolicy(\cdot | \ob_p, \hidden_p, \pweight)$ \Comment{Sample marginal tax rates; update planner hidden state}
        \Else
            \State $\texttt{no-op}, \hidden_p \Leftarrow \plannerpolicy(\cdot | \ob_p, \hidden_p, \pweight)$ \Comment{Only update planner hidden state}
        \EndIf
        \State $\st', \V{\ob}', \ob_p', \V{\rew}, \rew_p \Leftarrow \texttt{Env.step}(\st, \Ac, \mtaxrate)$  \Comment{Next state / observations, pre-tax reward, planner reward}
        \If{$t$ mod $\taxperiodlen$ = \taxperiodlen-1}  \Comment{Last timestep of tax period}
            \State $\st', \V{\ob}', \ob_p', \V{\rew}, \rew_p \Leftarrow \texttt{Env.tax}(\st', \mtaxrate)$  \Comment{Apply taxes; compute post-tax rewards}
        \EndIf
        \State $D \Leftarrow D \cup \{(\V{\ob}, \Ac, \V{\rew}, \V{\ob}')\}$ \Comment{Update agent transition buffer}
        \State $D_p \Leftarrow D_p \cup \{(\ob_p, \mtaxrate, \rew_p, \ob'_p)\}$ \Comment{Update planner transition buffer}
        \State $\st, \V{\ob}, \ob_p \Leftarrow \st', \V{\ob}', \ob'_p$
    \EndFor
    \State Update $\mweight, \pweight$ using data in $D, D_p$ and $\sA$.
    \State $D, D_p \Leftarrow \{\}, \{\}$ \Comment{Reset agent and planner transition buffers}
    \If{episode is completed}
        \State $\st, \V{\ob}, \ob_p, \V{\hidden}, \hidden_p \Leftarrow \st_0, \V{\ob}_0, \ob_{p, 0}, \V{\hidden}_0, \hidden_{p, 0}$ \Comment{Reset episode}
    \EndIf
    \If{criterion $C$ is met}
        \Return $\theta, \phi$
    \EndIf
    \EndWhile
\end{algorithmic}
\end{algorithm}

\section{Improved Social Outcomes with AI Agents}
\label{sec:expai}
\subsection{Baseline Methods}
\label{sect:baselines}
We now show empirically that the AI Economist can outperform baseline tax policies.
In particular, we will compare the following tax models:
\begin{itemize}
 \item free-market (no taxes),
 \item US federal single-filer 2018 tax schedule,
 \item Saez tax formula (adapted for a multi-period setting), and
 \item the AI Economist planner. %
\end{itemize}

The specific tax rates set by these models are depicted in Figure \ref{fig:tax_rates_ai}.
See the related work (Section \ref{sec:relatedwork}) for a broader discussion on the various tax frameworks proposed in the optimal tax literature, including linear tax models and analytical approaches to dynamic taxation in sequential economies.\footnote{We have also conducted experiments with linear planner models $\tax(\st) = \brcka{\V{\weight}, \st_{nonspatial}}$, but found they significantly underperform compared to all non-trivial tax models mentioned above.
Furthermore, we found that pure income redistribution leads to close-to-perfect equality, but very low productivity levels, and as a result, significantly worse social metrics.
As such, we do not include results for these models.}

All tax models set tax rates for a bracketed tax schedule, and use the same income brackets, following the 2018 US federal income tax schedule and scaling so that USD 1000 corresponds to 1 Coin:
\eq{
 \V{\bracketcutoff} &= [0, 9700, 39475, 84200, 160725, 204100, 510300, \infty] \quad\textrm{(USD)} \\
 &= [0, 9.7, 39.475, 84.2, 160.725, 204.100, 510.3, \infty] \quad\textrm{(Coin)}.
}

\paragraph{US Federal Income Tax Rates (Single-filer, 2018).}
The bracket tax rates are given by:
\eq{
 \V{\mtaxrate} = [0.1, 0.12, 0.22, 0.24, 0.32, 0.35, 0.37].
}

\paragraph{Saez Tax Formula (single-period).}
A prominent analytical treatment of optimal taxation is given by
\cite{saez_using_2001}, who proposes a closed-form solution for optimal tax rates in a single-period economy.

Let $\pdf$ and $\cdf$ denote the probability density and cumulative density function on income, respectively.
The Saez framework assumes the planner can observe the population's distribution over incomes $\income\sim\pdf(\income)$.
Here, $\income$ and the associated density functions refer to \textit{pre-tax} income within a single tax period.

\cite{saez_using_2001} works with the linear-weighted family of social welfare functions (Eq~\ref{eq:inv_income_weighted_utility}), and defines the {\em social marginal welfare weights} as
\begin{align}\socialmarginalwelfareweight_i = \frac{d\socialwelfare}{d\utility_i}\frac{d\utility_i}{d\money_i} = \socialwelfareweight_i \frac{d\utility_i}{d\money_i}.
\end{align}

Weight $\socialmarginalwelfareweight_i$ represents the change in social welfare due to a change in agent $i$'s endowment.\footnote{In the optimal tax theory literature the derivative of utility is taken with regard to an agent's consumption, which reflects its available money after taxes and redistribution. Endowment plays the same role as consumption in our model.}
The weights $\socialwelfareweight_i$, and implied social marginal welfare weights, $\socialmarginalwelfareweight_i$, parameterize the planner's objective, and encode a social choice, for example emphasizing agents with low wealth over agents with high wealth.
In instantiating Saez's framework, %
one available choice is to treat these social marginal welfare weights as the primitives in the model.
We do this, and  set the social marginal welfare  weights for the purpose of applying Saez's framework to be $\socialmarginalwelfareweight_i = \frac{1}{\income_i}$, also normalizing weights so that  $\sum_{i\in\mathcal{I}}\socialmarginalwelfareweight_i = 1$ (see also Section \ref{sect:socialwelfarefunctions}).
This  framework  does not explicitly optimize for the product of
equality and productivity. However, we  find empirically that optimizing with this choice for the social marginal welfare
weights tends to improve the product of equality and productivity.

To define the Saez framework, let $\alpha(\income)$ denote the {\em marginal average income at income $\income$, normalized by the fraction of incomes above $\income$}, i.e.,
\begin{align}
\alpha(\income) &= \frac{\income \cdot \pdf(\income)}{1-\cdf(\income)}.
\end{align}
Let $G(\income)$ denote the {\em normalized, reverse cumulative Pareto weight over incomes above a threshold $\income$}, i.e.,
\begin{align}
 G(\income) = \frac{1}{P(\income'\geq \income)} \int_{\income'=\income}^{\infty} p(\income') g(\income') \mathrm{d}\income'.
\end{align}
 where $g(\income)$ is the normalized social marginal welfare weight of an agent earning income $\income$.
In this way, $G(\income)$ represents how much the social welfare function weights the income  above  threshold $\income$.
Let {\em elasticity} $\elas(\income)$ denote the {\em average sensitivity of an agent's income
to changes in the tax rate}, defined as
\begin{align}
 \elas(\income) = \frac{\mathrm{d}\income / \income }{\mathrm{d}(1-\mtaxrate(\income)) / (1-\mtaxrate(\income))}.
\end{align}

\cite{saez_using_2001} shows that the optimal marginal tax-rate at pre-tax income $z$
is
\eq{
 \mtaxrate(\income) &= \frac{1-G(\income)}{1-G(\income) + \alpha(\income) \elas(\income)} \label{eq:saezformula}.
}

The salient property of this formula is that it does not depend on the agent's utility function, but rather depends  on the population's income distribution, $\pdf(\income)$, this  defining $\alpha(\income)$ and $G(\income)$, and the tax elasticity of income, $\elas(\income)$. Both of these quantities are, at least in principle, measurable.
In practice, a  challenge in applying the Saez formula is in estimating the tax elasticity of income, which is highly non-trivial in real-world economies. See \cite{gruber2002elasticity} for an extensive review of empirical approaches for the Saez framework.

The resulting tax schedule depends sharply on the shape of the income distribution. A log-normal-like income distribution, for example, leads to regressive taxes, with lower marginal rates at higher incomes, while a Pareto-like distribution leads to progressive taxes,  with higher marginal rates at higher incomes (see~\cite{mankiw_optimal_2009}).

\paragraph{Saez Tax Formula (multi-period).}
In our experiments, we apply the Saez formula to the multi-period setting by estimating the tax elasticity of income at the start of each tax period, and then appealing to Eq~\ref{eq:saezformula}. For this, we make use of a buffer $D = \brckcur{\brck{\income_{i_\alpha}, \mtaxrate_{i_\alpha}}}_{\alpha}$, which is a set of pairs of observed incomes and tax rates
in a window of previous tax periods, where the index $\alpha$ refers to a datapoint coming from agent $i_\alpha$.

Following \cite{gruber2002elasticity}, we assume constant tax elasticity $\tilde{e}$, with
\eq{\label{eq:elasticitysimpleformula}
\income_t = \income^0 \cdot (1-\mtaxrate_t)^{\tilde{e}}.
}
Hence, we can write:
\eq{\label{eq:elasticitysimpleformula_for_ols}
\log (\income_t) = \tilde{e} \cdot \log (1-\mtaxrate_t) + \log ( \income^0),
}
where $\income^0$ is the income that would result from zero taxes. Given the buffer $D$
collected from multiple rollouts, we estimate $\tilde{e}$ using ordinary least-squares regression on Eq~\ref{eq:elasticitysimpleformula_for_ols}.
In particular, we make use of the 30,000 most recent incomes and tax rates observed during rollout episodes, and find that this leads to stable estimates for the average elasticity $\tilde{e}$.

\paragraph{AI Economist.}
For the AI Economist, we make use of a deep neural network to set the marginal tax rate in each bracket, denoted
\eq{
 \mtaxrate \sim \plannerpolicy(\ob^\text{world}_{p, t}, \ob^\text{agent}_{p, t}, \ob^\text{market}_{p, t}, \ob^\text{tax}_{p, t}, \hidden_{p, t-1}; \pweight).
}

This shares the same general organization as the agents' policy model (Section \ref{sec:agent_utility}).
Indeed, the planner and agent policy networks use the same basic network architecture (Figure \ref{fig:neural_network_policies}).
However, the information in the planner observations differs from that in agents in some important ways.
For instance, the planner observes the full spatial state of the world in $\ob^\text{world}_{p, t}$, and the planner observes all agents' public states in $\ob^\text{agent}_{p, t}$ but does not observe any of their private states, observing endowments but not skills.
Section \ref{sec:details_of_environment} of the Appendix offers a detailed explanation of the different observations available to the agents and the planner.

\subsection{Training Strategy: Two-phase Training and Tax Curricula}
\label{sec:trainingstrategy}
As discussed in Section \ref{sect:innerouterrl}, the joint optimization problem posed by the inner-outer RL approach can lead to instability during learning.
One source of instability is that high tax rates cause large income penalties  during training, even for actions that might be optimal under low tax rates. Effective agent behaviors can be hard to learn due to this kind of noisy  feedback from an unconstrained, suboptimal planner  that generates random tax rates. We have found this to be especially problematic in the initial  phases of learning.

To stabilize learning we use a {\em two-phase training approach}.
In the first phase, we train a collection of agent models for a set of random seeds and without any taxes applied (the free-market scenario). This results in a set of agent models (one for each random seed) that are well adapted to the general game dynamics.

In phase two, we resume training, but with one of the studied tax models active.
In the case of the AI Economist, we also allow the planner to continue to adapt, along with  continued agent learning.
To avoid unstable learning dynamics created by the sudden introduction of taxes, we impose an {\em annealing schedule} over the early portion of phase two, during which a maximum limit on the allowable, marginal tax rates is linearly annealed from 10\% to 100\%.

Furthermore, we find that  {\em  entropy regularization} of the planner policy is necessary to achieve good outcomes in the face of these complex, joint learning dynamics.
Entropy regularization adds the policy's entropy as an additional, weighted term in the policy gradient objective, and is defined as
\eq{ \texttt{entropy}(\policy) = -\E_{\ac\sim\policy(.|\st)}\brcksq{\log\policy(\ac | \st)}.}

The use of this entropy term promotes  policies that explore more when used together with on-policy learning, which samples   trajectories according to the current policy $\policy$ \citep{williams1991function,mnih2016asynchronous}.

We perform experiments using the RLlib framework~ \citep{Liang2018_RLlib}.
We use {\em proximal policy gradients}~\citep{schulman2017proximal} and the Adam optimizer~ \citep{kingma_adam:_2014} to compute policy gradients.
Samples were collected from 60 environment replicas in parallel, using a sampling horizon of 200 timesteps between policy update iterations (a full episode consists of 1000 timesteps).
Trajectories were chunked into subsequences of length 50 for training the recurrent networks.
For more details, see the Appendix.
All experiments performed phase two training with 400 million samples, which we found to be sufficient for both agent and planner models to converge to stable policies.
The annealing schedule allows the maximum  marginal tax rate to reach 100\% by 54 million samples.
\begin{figure}[t!]
  \begin{small}
    \begin{center}
    \includegraphics[width=0.95\textwidth]{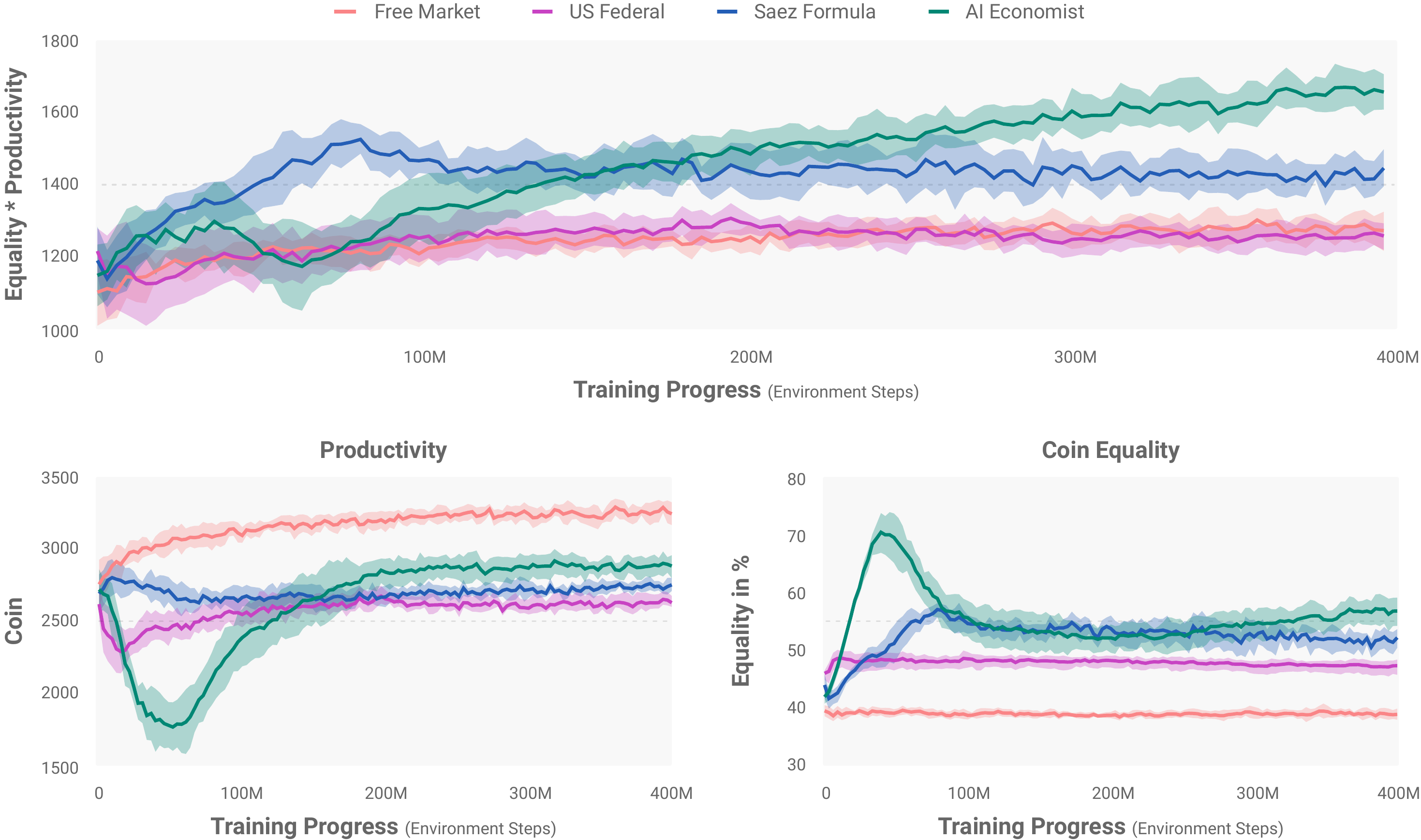}
    \end{center}
    \caption{Empirical training progress for all models. The AI Economist (Green) achieves significantly better social outcomes than the baseline models. All baseline models have converged.}
    \label{fig:wandbcurves}
  \end{small}
\end{figure}

\subsection{Equality, Productivity, and Social Welfare Metrics}
\begin{figure}[t!]
  \begin{small}
    \begin{center}
    \includegraphics[width=0.617\textwidth]{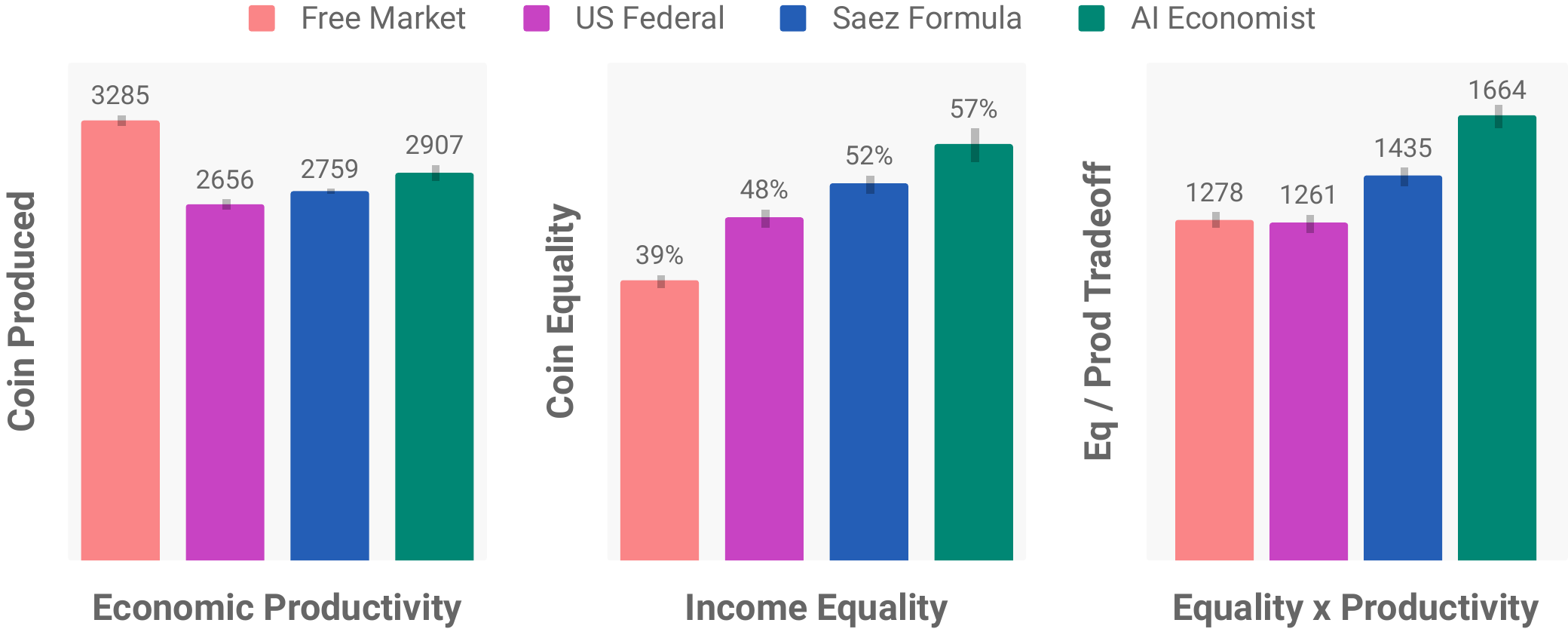}
    \end{center}
    \caption{Comparison of overall economic outcomes (error bars show the variance during the final 20 million training steps, across 10 random seeds). The AI Economist achieves significantly better equality-productivity trade-offs compared to the baseline models. All baseline models have converged.}
    \label{fig:prod_eq_welfare_ai}
  \end{small}
\end{figure}
We compare economic outcomes under the AI Economist with the free market (no taxation or redistribution), a simulated US Federal tax schedule, and the tax policy that results from the Saez framework~\citep{saez_using_2001}.

For all four treatments we use reinforcement learning to optimize the behavior of the economic AI agents. The results are shown in Figure \ref{fig:prod_eq_welfare_ai}. Productivity (left panel, higher is better) measures the total amount of income generated within an episode (analogous to GDP). Taxation always results in a decrease of productivity when compared with the free market, but the loss in productivity is the smallest under the AI Economist. Income equality (middle panel, higher is better), which is defined as 1 - Gini index and computed at the end of an episode (higher Gini index means incomes are less equal), is highest under the AI Economist.
The product of equality and productivity (right panel, higher is better) measures the balance between equality and productivity. The AI Economist achieves a 16\% gain improvement over the next best model, which is the Saez model. The AI Economist also improves equality by 47\% compared to the free-market, at only an 11\% decrease in productivity.

As discussed in Section \ref{sec:opttax}, the challenge in setting taxes stems from the inherent trade-off between equality and productivity.  This can be  seen in the empirical results, where
redistribution  improves equality but at the cost of productivity.\footnote{We also find in our experiments that total redistribution (such that all workers have the same income after redistribution) yields perfectly equal but highly unproductive economies and very low equality-vs-productivity trade-offs.}
This property naturally emerges  in our simulations, by allowing agents to learn optimal responses to  taxes.

In summary, these results demonstrate (1) that our framework allows us to reproduce the central challenge considered in optimal taxation theory, the trade-off between equality and productivity, (2) that the severity of this trade-off depends on the choice of tax schedule, and (3) that RL can be used to optimize tax policies.

\subsection{Tax Schedules and Wealth Redistribution after Taxes and Subsidies}
\begin{figure}[t!]
  \begin{minipage}[t]{0.48\linewidth}
    \begin{small}
    \begin{center}
      \includegraphics[width=\textwidth]{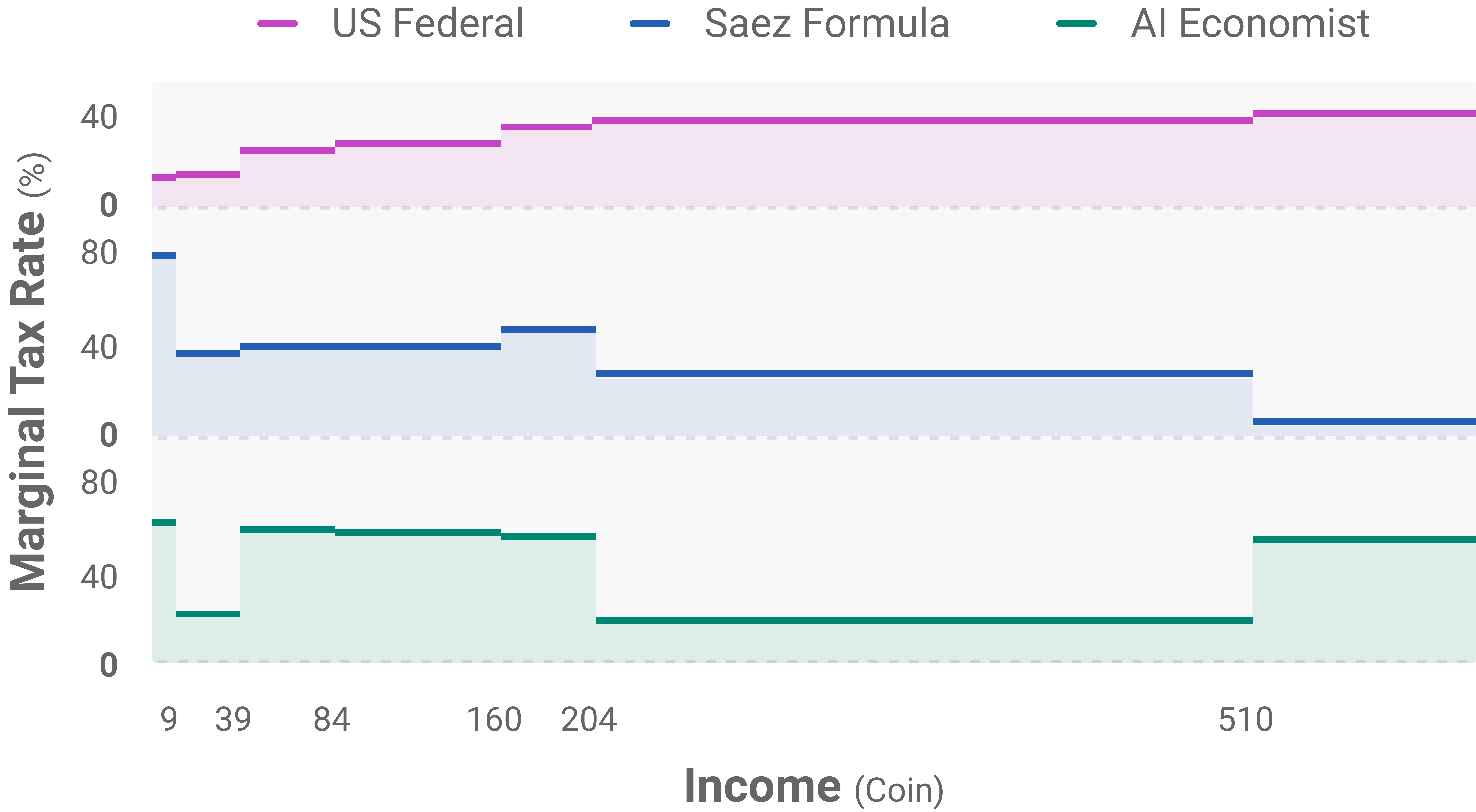}
    \end{center}
    \caption{Comparison of average tax rates per episode. Variances within the Saez and AI Economist schedules are not shown. On average, the AI Economist sets a  higher top tax rate than both of the US Federal and Saez tax schedules. The free-market collects zero taxes.
    }
    \label{fig:tax_rates_ai}
    \end{small}
  \end{minipage}
  \hspace{0.03\linewidth}
  \begin{minipage}[t]{0.48\linewidth}
    \begin{small}
    \begin{center}
      \includegraphics[width=\textwidth]{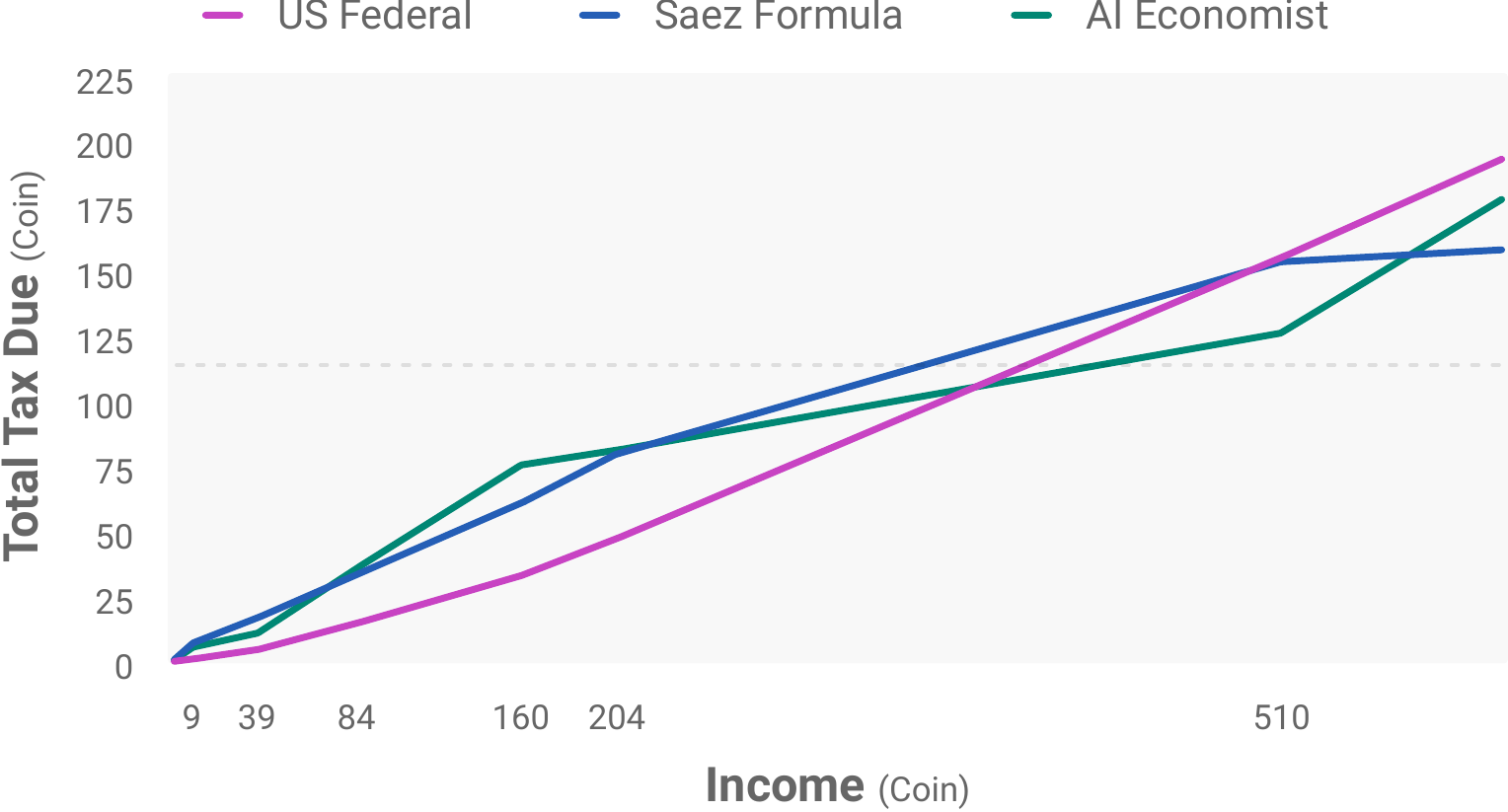}
    \end{center}
    \caption{The effective taxes payable as a function of income. The taxes grow faster under the Saez and AI Economist schedules than under the US Federal schedule. But this does not include the effect of subsidies that arise from this collection of taxes (in effect, lower incomes  receive net subsidies, see Figure \ref{fig:tax_impact_per_agent}).}
    \label{fig:effective_tax_ai}
    \end{small}
  \end{minipage}
\end{figure}
\begin{figure}[t!]
  \begin{small}
    \begin{center}
    \includegraphics[width=0.8\textwidth]{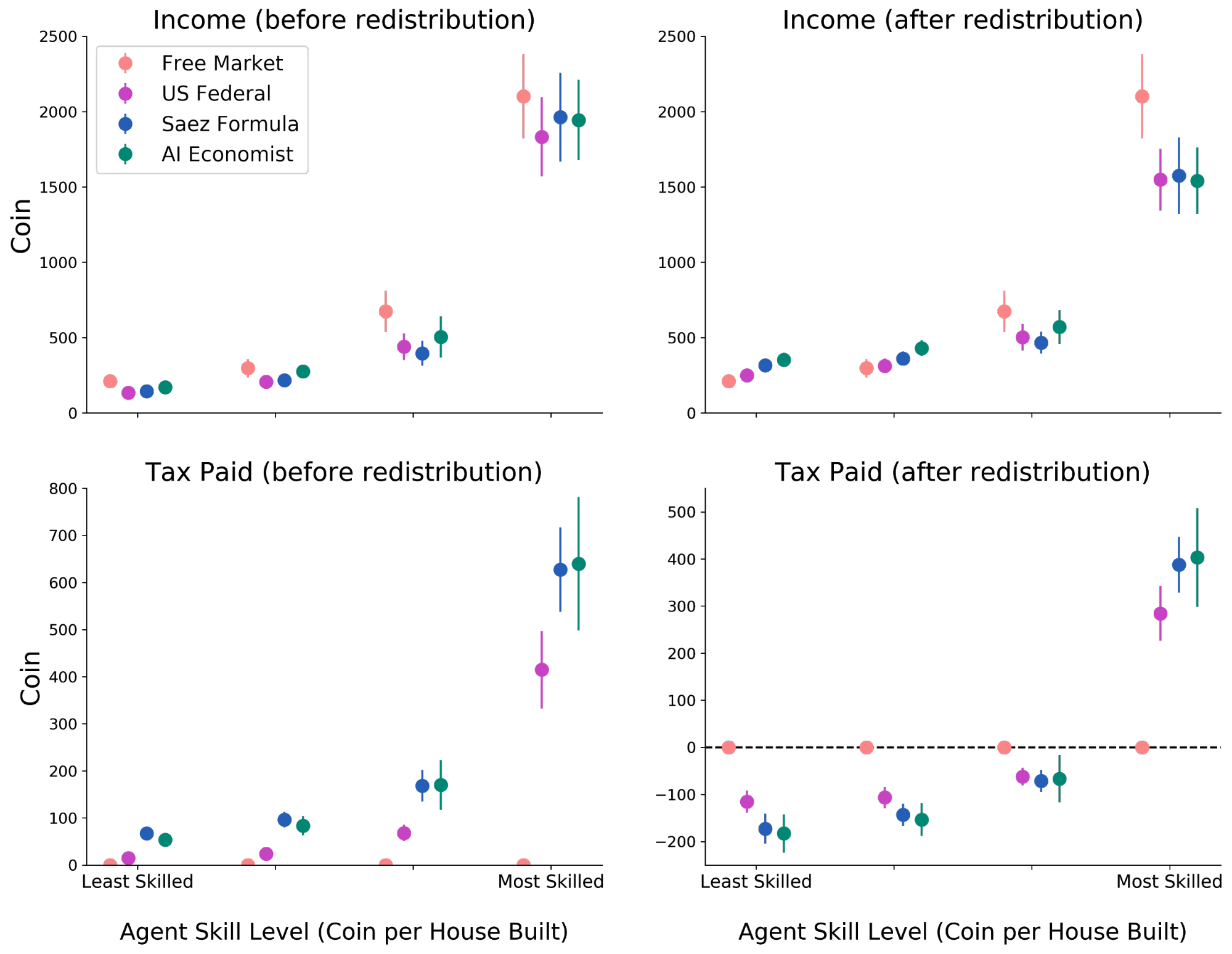}
    \end{center}
    \caption{Agent-by-agent averages after sorting by skill. Income before redistribution (top-left) shows the average pre-tax income earned by each kind of agent. The amount of tax paid before distribution is shown in the bottom left. The amount of tax paid after redistribution is shown in the bottom right (the lower skill agents receive a net subsidy).   The income after redistribution (top-right) shows the net average coin per agent at the end of the episode (the lower-skilled agents have higher net income under the AI Economist’s tax scheme).}
    \label{fig:tax_impact_per_agent}
  \end{small}
\end{figure}
\paragraph{Comparing Tax Schedules.}
All tax models control the marginal tax rates applied to each of seven income brackets (see Figure \ref{fig:tax_rates_ai}, which illustrates the average bracket rate set by each model).
We set up the economic simulation such that the fraction of agent incomes per income bracket are in rough alignment with those in the US economy\footnote{Based on preliminary experiments with the US Federal tax policy.}.

The 2018 US Federal tax rates are progressive, with a marginal tax rate  that increases with higher  income. For the present setting, and with the social welfare objective that we adopt, the Saez tax framework mostly sets a regressive tax schedule, with a marginal tax rate that decreases with higher income. The AI Economist features a more idiosyncratic structure, with a blend of progressive and regressive tax schedules. In particular, it sets a higher top tax rate (on income above 510), a lower tax rate for incomes between 160 and 510, and both higher and lower tax rates on incomes below 160.
\paragraph{Effective Tax After Redistribution.}
The AI Economist's tax schedule provides higher subsidies to low income agents than the baselines. %
The  agents have different skill levels, and the learned behaviors, incomes, and amount of tax paid all depend heavily on skill. Figure \ref{fig:tax_impact_per_agent} presents the agent-by-agent averages after sorting by skill. Income before redistribution (top-left) shows the average pre-tax income earned by each kind of agent. Tax paid is shown in the bottom left. The effect of redistribution,  which equally divides collected taxes among the agents, is that the lower-skilled agents receive a net subsidy (bottom right). The income after redistribution shows the net average coin per agent at the end of the episode (top right). The lower-skilled agents have higher net income under the AI Economist than under the other models.

\paragraph{The Impact of Tax on Economic Activity.}
\begin{figure}[t!]
  \begin{small}
    \begin{center}
    \includegraphics[width=0.9\textwidth]{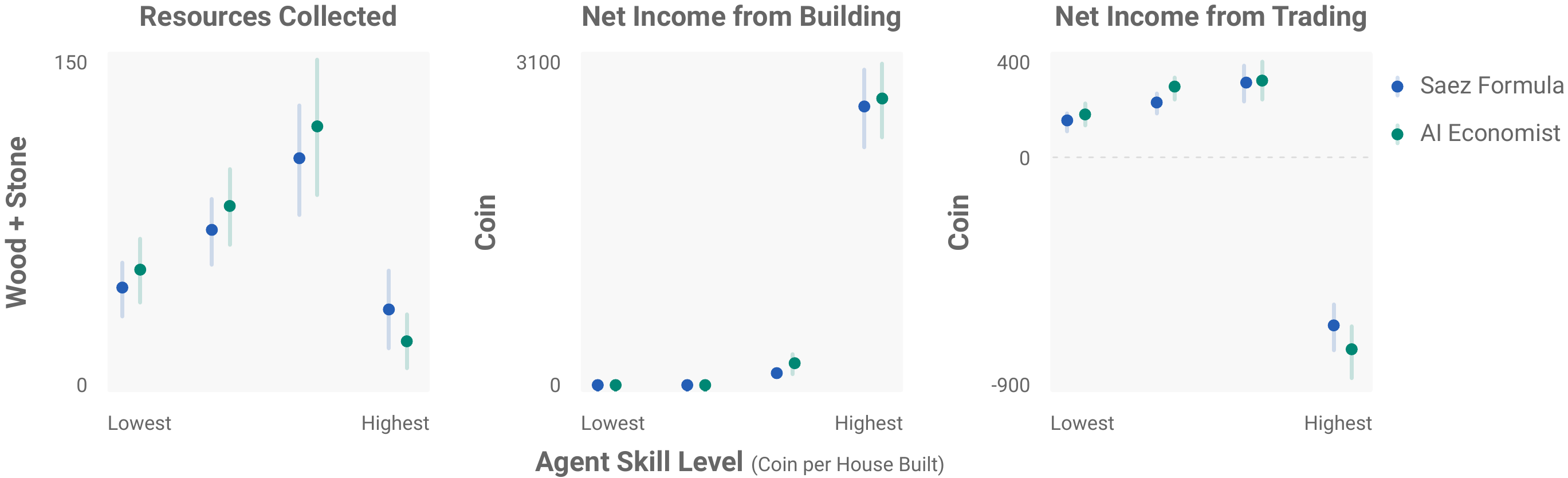}
    \end{center}
    \caption{
    Comparing the impact of tax on economic activity for the Saez formula and the AI Economist.   Left: Average number of resources (both wood and stone) collected per episode for each of the four agents.
      Middle: Average (per episode) total income earned from building.
      Right: Average (per episode) total income earned from trading. Negative income means the agent spends more than it earned.
      Lines indicate the standard deviation.
    }
    \label{fig:saez-aieconomist-comparison}
  \end{small}
\end{figure}

To form a better understanding of how taxes set by the AI Economist improve over those set by the Saez formula, which provides the strongest baseline, we compare their respective impacts on the agents' economic activity (Figure \ref{fig:saez-aieconomist-comparison}).
In both cases, the low-income agents  choose to specialize as ``gatherer-and-seller” agents.
Interestingly, these agents collect fewer resources under the Saez policy, and the high-skilled ``buyer-and-builder” agent compensates by increasing its own resource collection (Left panel).
This de-specialization  contributes to the decreased income generated through building under the Saez taxes (Middle panel), with this decrease accounting for  weaker productivity.

Because the Saez formula leads to a more regressive tax structure than the AI Economist, the latter yields higher equality through \textit{mechanical} effects (i.e. stronger redistribution).
Interestingly, the AI Economist also improves equality through \textit{behavioral} effects.
Under the Saez scheme, the ``buyer-and-builder'' collects more resources directly from the environment, meaning it makes fewer purchases from the other agents. Trading  serves to redistribute the income achieved from building houses to the ``gatherer-and-seller'' agents, but, owing to the behavioral differences, this redistributive effect is stronger under the AI Economist (Right panel).
From this perspective, the tax scheme discovered by the AI Economist appears better adapted to the complex economic interactions that shape both equality and productivity.

\subsection{Tax-Gaming Strategies}
\begin{figure}[t!]
  \begin{small}
    \begin{center}
    \includegraphics[width=0.7\textwidth]{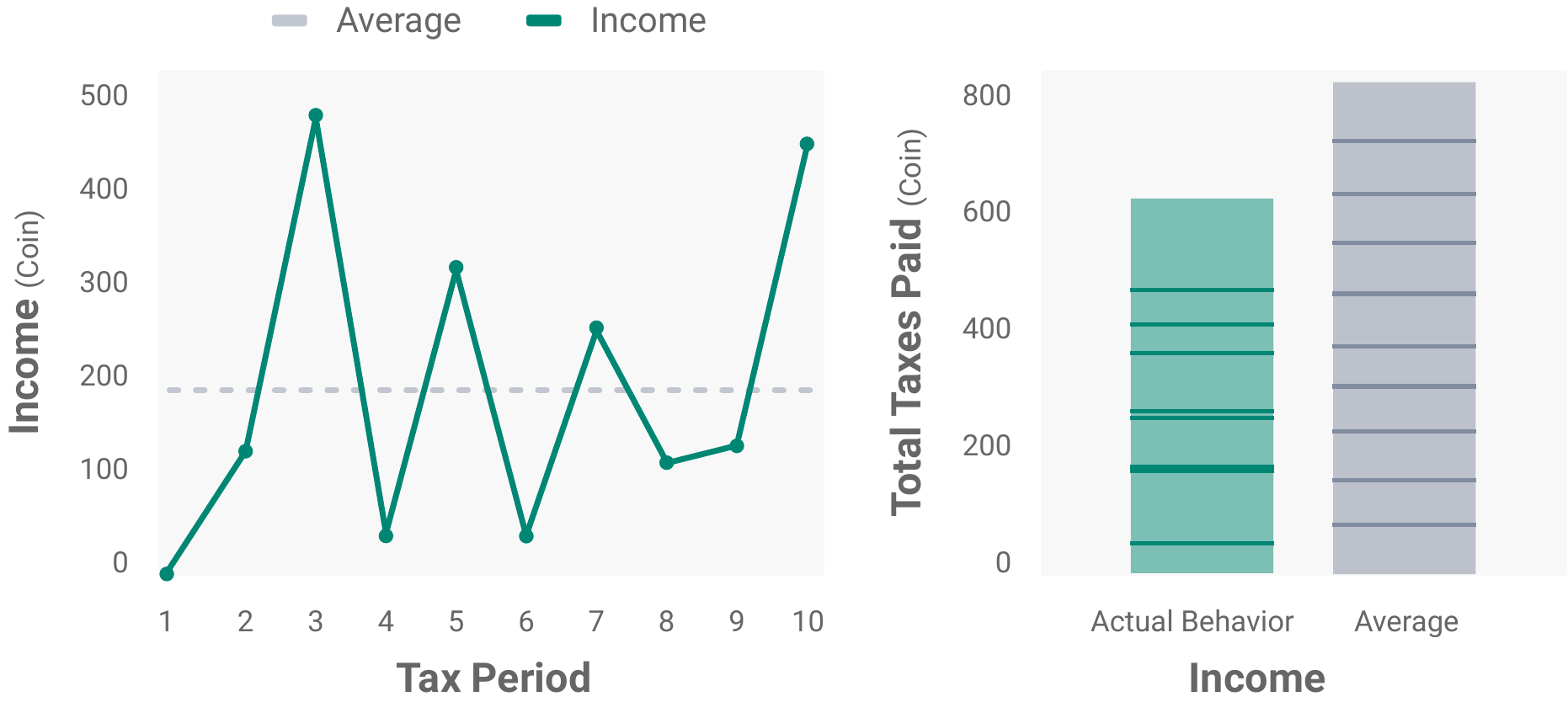}
    \end{center}
    \caption{Left: Income of the highest-skilled agent for each tax period in an example episode with the AI Economist (green line).  The dashed grey line shows the agent's average income. Right: Comparison of the total amount of tax the agent owed based on its actual  income (green) and the tax it would have owed if it reported its average income in each period (gray). Each box in the column denotes the tax obligation in a single period.
    }
    \label{fig:tax_episode}
  \end{small}
\end{figure}

Figure~\ref{fig:tax_episode} provides an example of the  income and taxes collected during an episode of the AI Economist environment, shown here after the tax policy has converged. Recall that each episode is divided into ten tax periods of equal length. At the start of each period, a new tax schedule is set by the AI Economist according to the new world state. Agents act in the environment,  earn income, and  taxes are collected at the end of the period according to the tax schedule and redistributed.

We see emergent tax gaming, where AI agents  learn to lower their average effective tax by alternating between earning high and low incomes in each period, rather than smoothing their income across  tax periods.
Figure~\ref{fig:tax_episode} (left) shows there is considerable variability in income earned from period to period, shown here for the highest skilled agent.  Figure~\ref{fig:tax_episode} (right)
 shows  the total amount of taxes paid given this behavior, together with the total taxes that would be paid if the  income had been smoothed across periods.

We see this kind of tax avoidance behavior in our experiments  for both the Saez and AI Economist models, which   feature lower top tax rates (regressive schedules), making it more tax-efficient to earn high incomes. This underscores the richness of the simulation-based learning framework. Moreover, the  AI Economist remains effective even in the face of this kind of strategic  behavior.

\section{Improved Social Outcomes with Human Participants}
\label{sec:exphuman}
We have also explored whether AI-learned tax policies improve social outcomes in economic simulations with human participants who earn real money.
To do so, we conducted experiments on the Amazon Mechanical Turk (MTurk) platform, with participants based in the US.
We find that the AI Economist tax policy can  transfer to simulations with people  without extensive recalibration or fine-tuning. The AI Economist achieves equality-productivity trade-offs that are competitive with the strongest baseline, the Saez tax policy (Equation \ref{eq:saezformula}), and achieves higher inverse income-weighted social welfare.

\subsection{Experimental Methodology}
\paragraph{Simulation Environment for Human Participants.}
We used the same world layout as in the AI experiments. The world map features four quadrants, mostly separated from each other by water. Each quadrant contains only stone, wood, both resources, or neither resource.
Each participant controls an agent with a fixed skill, set as the mean of the quartiles of a Pareto distribution with exponent $a=4$ and scale $m=1$, and each starting in one of the four corners. This starting location was randomized for each episode.

\begin{figure}[t!]
 \begin{center}
  \begin{minipage}[b]{0.43\linewidth}
   \begin{small}
    \begin{center}
     \includegraphics[width=\linewidth]{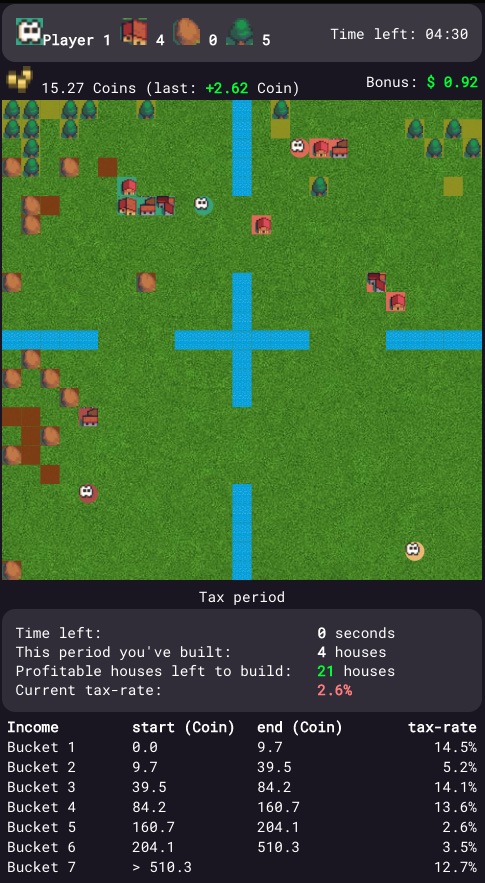}
    \end{center}
    \caption{The web graphical user interface that human participants used in experiments.}
    \label{fig:human_exp_gui}
   \end{small}
  \end{minipage}
  \hspace{0.01\linewidth}
  \begin{minipage}[b]{0.48\linewidth}
   \begin{small}
    \begin{center}
     \includegraphics[width=\textwidth]{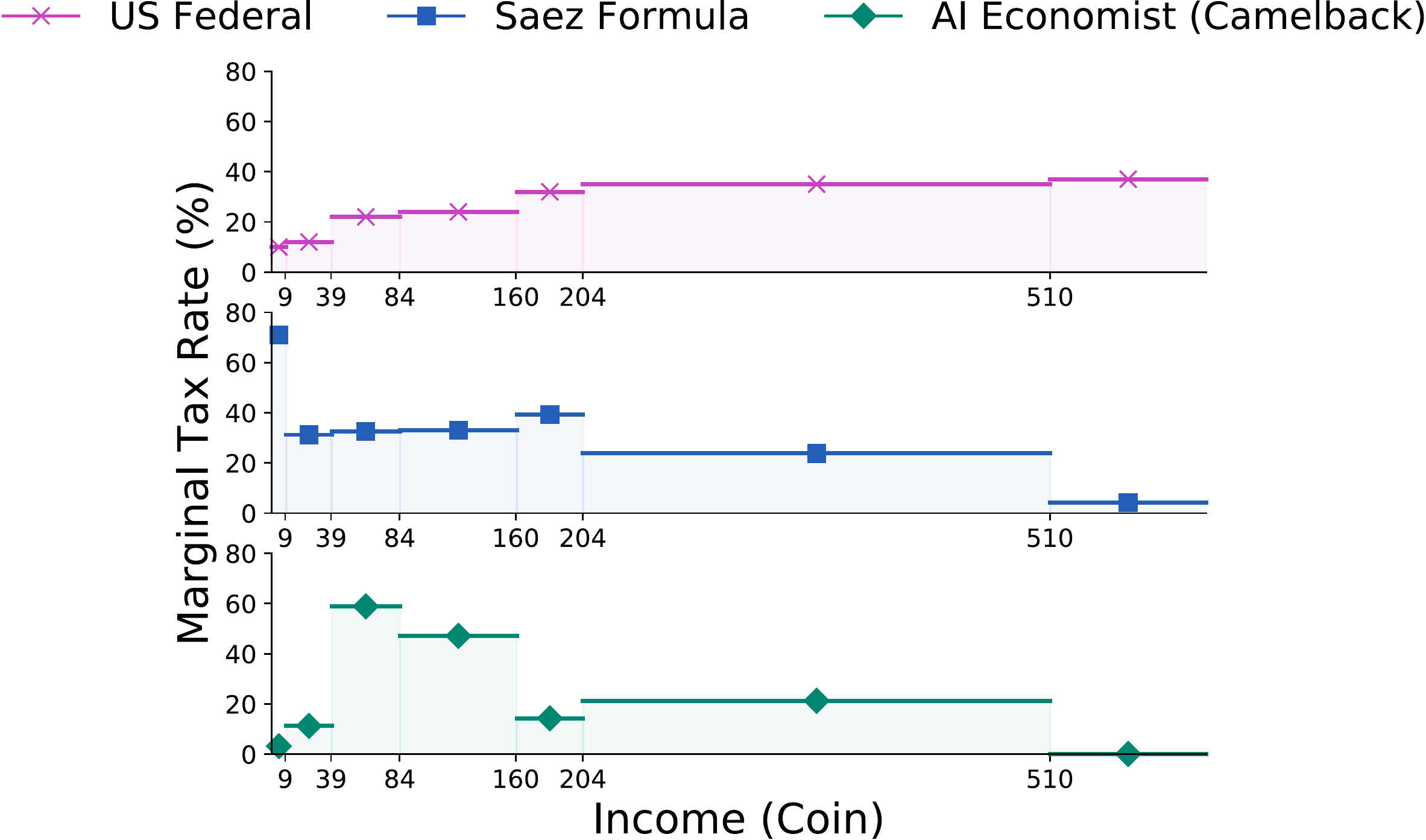}
    \end{center}
    \caption{The \camelback{} model used in experiments with human participants. It features higher tax rates for incomes between 39 and 160 Coins compared to baselines.}
    \label{fig:tax_rates_with_camelback}
    \begin{center}
     \includegraphics[width=\textwidth]{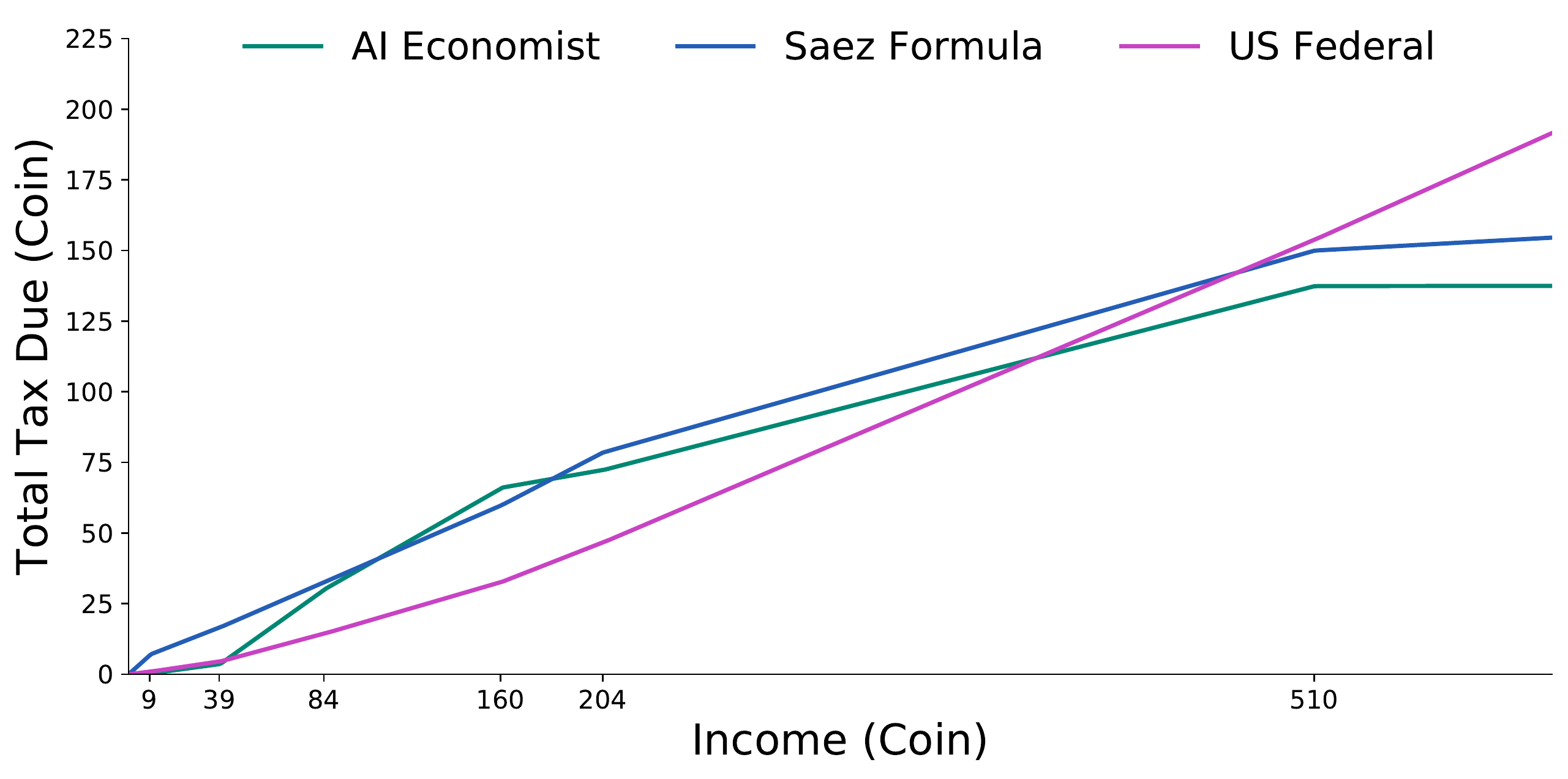}
    \end{center}
    \caption{The effective taxes payable as a function of income under the \camelback{} schedule. The taxes grow faster under the Saez and AI Economist schedules. Note that these do not include the effect of subsidies. In effect, lower income workers receive net subsidies.
    }
    \label{fig:effective_tax_camelback}
   \end{small}
  \end{minipage}
 \end{center}
\end{figure}

We make several modifications to account for human response times, allowing  for an acceptable experiment duration and simplified controls:
\begin{itemize}
 \item We disable trading. We experimented with several trade inferfaces, but found none that were usable enough. Even without trading, humans experienced the same economic drivers, namely utility-maximization and diminishing returns, as the AI agents.
 \item The only kind of action that is associated with a labor cost is the  build action. Moving around the environment and collecting resources has zero cost. To compensate for this, the cost of building a house is 50\% higher than in the AI experiment (15 vs.~10 labor units per house).
 \item Each episode lasts five minutes. To allow for acceptable human response times we set the frame rate to ten frames per second (each frame corresponds to a new world state).
 This provides participants with enough time to achieve reasonable performance, partially correcting for the lower response times compared with AIs.
 \item Each episode lasts 3000 timesteps rather than 1000 timesteps, with each tax period consisting of $300$ timesteps (keeping ten tax periods in each episode).
\end{itemize}
\paragraph{Graphical User Interface.}
We developed a web-based interface to let people operate in an economic simulation similar to the one used in the AI experiments. For a full visualization of the experimental flow and pages, see Appendix \ref{app:human_exp_details}.
The interface (Figure \ref{fig:human_exp_gui}) displays agents' endowments (coin, stone, wood, houses), the remaining episode time, the last change in coin endowment, the bonus in USD, the tax schedule, the current active tax rate (which depends on the income in the current tax period), and the remaining time in the current tax period.

We also provide participants with the number of profitable houses left to build (i.e., for how many more houses in the current tax period will it still remain profitable to build). This decision aid helped participants to better understand the economic environment, leading to less variance in the experimental results across trials.
Despite this guidance, participants frequently scored lower utility than in the AI experiments. Sometimes this would come about because of adversarial behavior of others, especially resulting from people blocking other people from accessing areas with resources, or finding ways to trap people in corners.
\paragraph{Zero-shot Transfer of Tax Models.}
The tax models  were transfered from the AI-only setting.
The US Federal tax rates were unchanged.
For the Saez model, we used the average tax rate observed during an episode once training has converged.
For the AI Economist, we identified an effective AI-driven tax schedule from the AI experiments conducted with low planner policy entropy regularization.\footnote{This model was chosen from a set of AI models that performed as well as or better than the baselines tax models.
In particular, we found that planner policies with high entropy did not generalize as well in the zero-shot transfer setting.
We did not retrain or fine tune the AI tax model.}
The particular tax schedule that we use has a \camelback{} style shape, and is depicted in Figure~\ref{fig:tax_rates_with_camelback}.
The effective taxes after redistribution are shown in Figure \ref{fig:effective_tax_camelback}.
The \camelback{} policy achieves competitive equality-productivity and weighted social welfare (Equation \ref{eq:inv_income_weighted_utility}) in the AI-only simulations, compared to the Saez tax model.

The productivity was lower in experiments with people. This is due to suboptimal human behavior, as well as lower human response times compared to AI agents.
To ensure that all tax policies could still make use of the full range of tax brackets, we calibrated the income bracket cutoffs to approximately match the income bracket occupancy rates to those in the AI experiments, achieving this by scaling the income cutoffs down by a factor of three.
\paragraph{The Experimental Protocol.}
We ran all experiments with US-based participants on Amazon Mechanical Turk.
Participants performed HITs (Human Intelligence Task). Each HIT consists of a sequence of four episodes, with a tutorial before each episode, and a post-episode survey. For detailed descriptions and visualizations of the experiment modules, see the appendix.

HITs were announced  in batches of 40-60, where each unique participant could accept one assignment from each batch (but could perform more than one HIT across different batches).
Batches were sized so that all assignments in the batch were completed within two hours, accounting for participant availability.
Participants were instructed not to communicate with each other. Experiments were conducted during 10am-12pm and 7-10pm, Pacific Time.
All participants were grouped into groups of four. Each group went through a sequence of four episodes, with each episode corresponding to a different tax policy (free market, US federal, Saez, and AI), these applied in random order to control for learning effects.
\paragraph{Payment.}
Each participant received \$5 base pay and a variable bonus of at most \$10 for each HIT. The bonus was proportional to the utility achieved by the participant, reflecting the post-tax income and the labor cost at the end of each episode.
The US dollar (USD) bonus was computed as
\eq{
 \textrm{USD bonus} = \textrm{Utility} \times 0.06,
}
where utility was measured in units of Coins achieved by the participant in the episode.
The effect is an average payment per HIT of \$11.26.
Since the average duration of a HIT was approximately 30 minutes, the effective income (approximately \$20/hour)  is substantially above the US federal minimum wage (\$7.25/hour). As such, we believe that the stakes should be high enough to encourage participants to try to maximize their bonus and avoid behaviors that result in a decrease in utility.

We use a set of qualification HITs to build a pool of around 300 qualified participants who are familiar with the instructions and the simulation environment.
Qualification HITs used exactly the same rules and environment as in the main task, with the only exception being that no information about taxes was given. For instance, in the qualification, participants did not observe what the tax schedule was, nor an explanation in the tutorial as to how taxes were applied.
In the main task, participants completed a tutorial that explained that taxes affected the income gained per house, and how this impacted their utility (and payments for the HIT).
In an exit survey, participants were asked about their strategy, why they thought they won or lost, and what was confusing about the experiment.
\subsection{Results}
\paragraph{Experiment Data.}
We report results on two batches of experiments.
In the first batch, groups were formed with the participants who were available at the end of an episode. This allowed as many users as possible to complete four episodes (we found during qualification batches that some participants experienced technical issues that prevented them from completing four episodes in sequence).
In the second batch, each group of 4 workers was fixed during a sequence of 4 episodes with 4 different tax models.
The first batch consisted of 57 episodes with 58 participants.
The second batch consisted of 68 episodes with 57 participants.

Feedback from participants and manual inspection of movement patterns in rollouts suggested that there were episodes in which one or more participants suffered from connectivity issues (as evidenced by extreme lag or disconnections), did not move around the world, or in which there were other factors that severely affected proper participation.
As such, we dropped  episodes from the analysis in which the overall productivity was less than 1000 Coin. This excluded 6 out of 57 episodes from the first batch and 8 out of 68 episodes from the second batch for our analysis.
\paragraph{Statistical Analysis.}
For the first batch, we test whether the difference in the social welfare  between the tax models is statistically significant. In particular, for each participant $a$,
we compute the mean value $Z_{ai}$ of the social objective (e.g., $\equality\times\productivity$) under each tax model $i$.
We then perform a two-sided t-test for the alternate hypothesis $Y_{a;vw}\neq 0$, with $p = 0.05$. The data consists of the differences $\brckcur{Y_{a;vw} = Z_{av} - Z_{aw}}$ for each pair of tax models $v$ and $w$, for each participant $a$ that experiences both models.

For the second batch, where group consistency was enforced, we perform this test at the group level. For each group $g$, we compute the mean value $Z_{gi}$ of the social objective (e.g., $\equality\times\productivity$) under each tax model $i$.
We then use a two-sided t-test to test the alternate hypothesis $Y_{g;vw}\neq 0$ with $p = 0.05$ on the set of differences $\brckcur{Y_{g;vw} = Z_{gv} - Z_{gw}}$, for each pair of tax models $v$ and $w$.
\begin{figure}[t!]
 \begin{center}
 \begin{minipage}[t]{0.6\linewidth}
  \begin{small}
   \begin{center}
    \includegraphics[width=\linewidth]{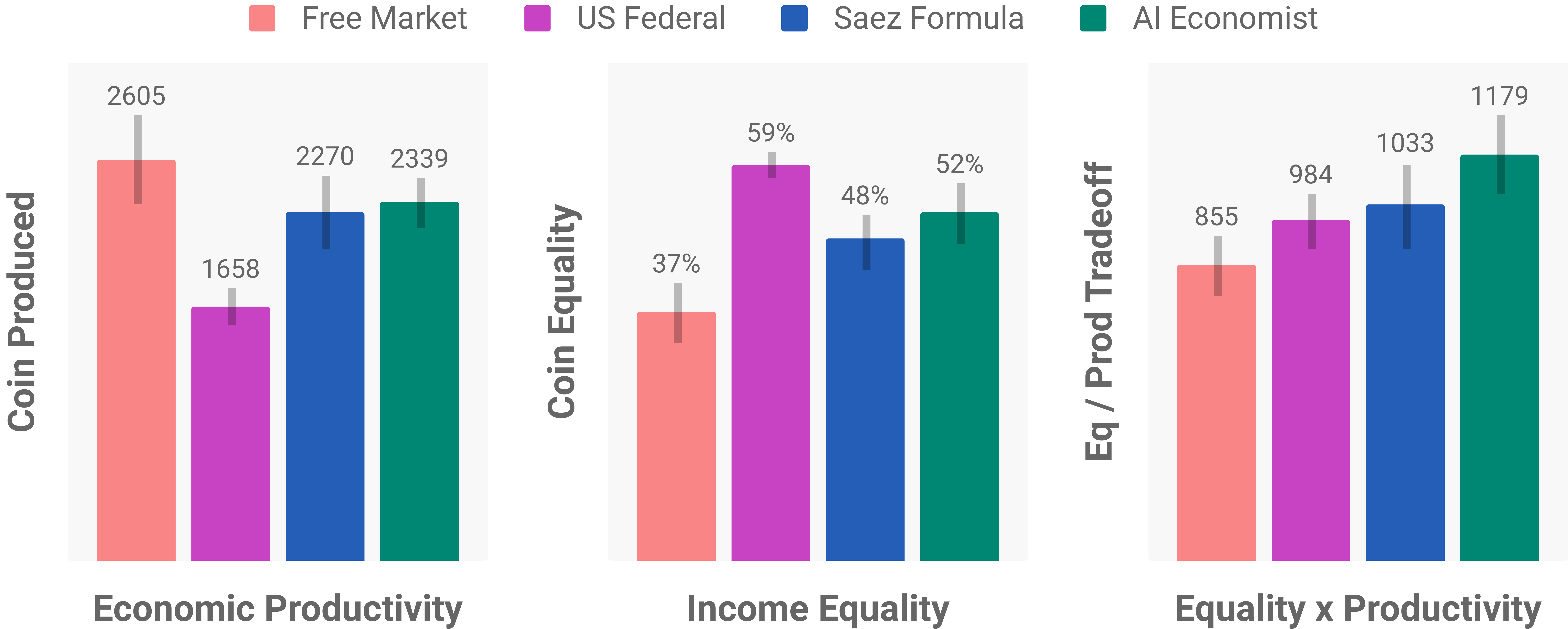}
   \end{center}
   \caption{Social outcomes with 58 human participants in 51 episodes (first batch episodes with productivity of at least 1000 Coin). Each episode involves four participants. The AI Economist achieves competitive equality-productivity trade-offs with Saez and US Federal, and statistically significantly outperforms the free market (at $p=0.05$). These results suggest a similar trend of improvement in equality-productivity trade-off as in the AI experiments.}
   \label{fig:eq_prod_human}
  \end{small}
 \end{minipage}
 \hspace{0.01\linewidth}
 \begin{minipage}[t]{0.375\linewidth}
  \begin{small}
   \begin{center}
    \includegraphics[width=\linewidth]{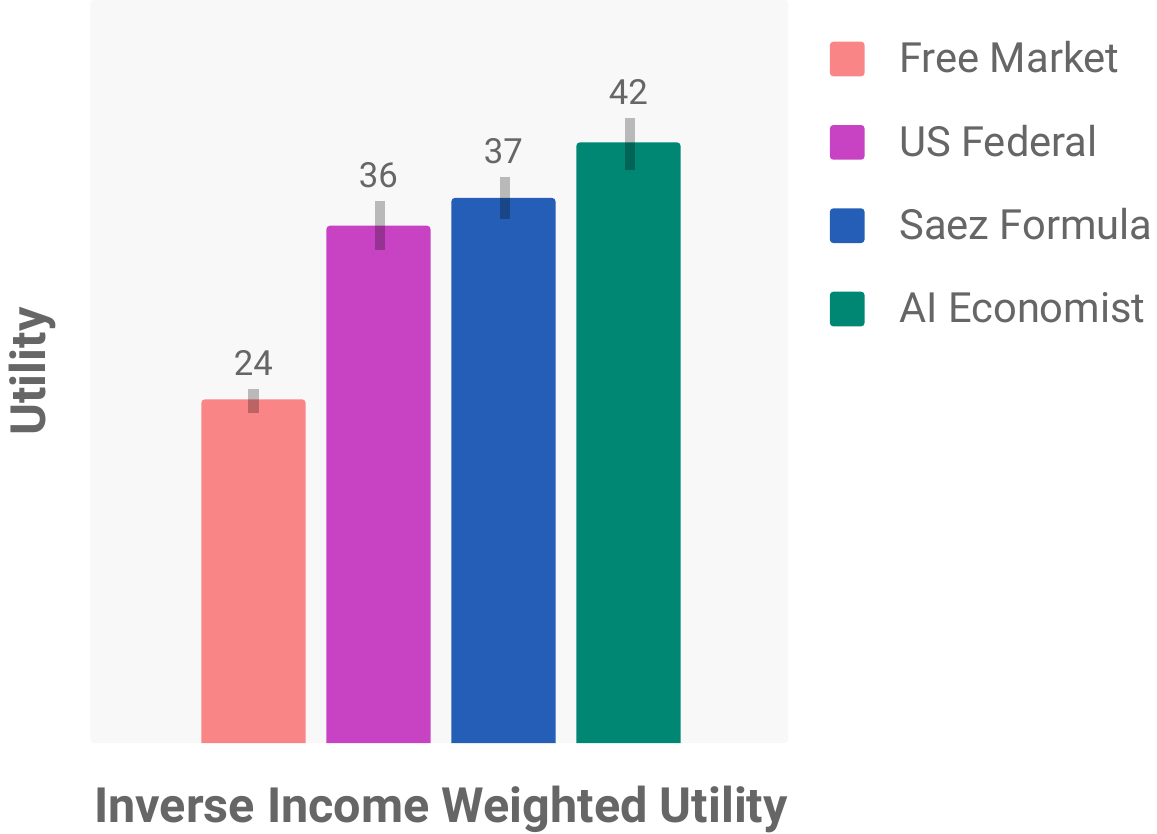}
   \end{center}
   \caption{Weighted average social welfare with 51 human participants in 60 episodes (second batch episodes with productivity of at least 1000 Coin). Each episode involves four participants. The AI Economist achieves significantly higher weighted social welfare than all baselines (statistically significant at $p=0.05$).
}
   \label{fig:inv_income_weighted_utility_humans}
  \end{small}
 \end{minipage}
 \end{center}
\end{figure}
\paragraph{Improved Social Outcomes.}
In experiments with human participants, the \camelback{} tax schedule achieves an equality-productivity trade-off that is comparable to the Saez model, and with better equality-productivity performance than the US Federal and free-market approaches (see Figure \ref{fig:eq_prod_human}). We observed large variance in productivity across episodes, which can be attributed to adversarial behavior and other factors that we discuss below.

We also evaluate the social welfare at the end of an episode,
using inverse post-tax endowments as social welfare weights:
\eq{\label{eq:omega_human_exp}
\socialwelfare_\eplen(\Money_\eplen,\vlabor_\eplen) = \sum_{i=1}^N \socialwelfareweight_i \cdot \util_i\brck{ \money_{i,\eplen}, \labor_{i,\eplen} },\quad
\socialwelfareweight_i = \fr{\tilde{\socialwelfareweight}_i }{ \sum_j \tilde{\socialwelfareweight}_j },\quad
\tilde{\socialwelfareweight}_i = \fr{1}{\money_{i,H}},
}
where $\money_{i, H}$ is the post-tax endowment of agent $i$ at the end of the episode of length $\eplen$, and $\socialwelfareweight$ is normalized such that $\sum_i \socialwelfareweight_i = 1$.
This evaluation objective
places more weight on agents with lower endowments  than those with higher endowments, considering
agent endowments at the end of an episode, and thus the cumulative effect of tax
policy over a sequence of ten tax periods.\footnote{This objective is related  to the choice we make about the tax policy objective  when  instantiating the Saez  framework, while deviating in a couple of important ways. First, the Saez framework  considers economies with a single tax period and does not consider the effect of taxation policy on the cumulative endowment.
Second, the particular choice we make in regard to the social marginal welfare weights, $\socialmarginalwelfareweight_i = \frac{1}{\income_i}$, when using Saez's framework to derive an optimal tax policy, does not  correspond directly to even the single period version of this inverse-income weighted objective, since by setting $\socialmarginalwelfareweight_i = \frac{1}{\income_i}$ it is as if  $\frac{d\utility_i}{d\money_i}=1$, and thus as if an agent's utility function is linear.}

The results of the experiment with respect to this objective are shown in Figure \ref{fig:inv_income_weighted_utility_humans}.
We can see that the \camelback{} tax schedule  significantly outperforms all baselines for this social
welfare objective.

Overall, the  relative  performance of the AI Economist compared with the various baselines
is similar for the experiments with AI agents and the experiments with human participants. In particular, even though the \camelback{} tax schedule is qualitatively different than the tax schedule that results from the Saez framework, it yields a competitive equality-productivity tradeoff in comparison with the schedule coming from the Saez model.
\subsection{Discussion}
The experiments with human participants are conducted in a zero-shot learning transfer setting, and the AI Economist performs well, even though there are a number of differences between the two settings.
Besides the modifications to the environments, other factors affecting the transfer from the AI environment to the human environment include:
\begin{itemize}
\item {\em AI and human behavior differs substantially}. For example, we have observed that humans display a higher frequency of adversarial behavior, such as blocking other people. These kinds of behaviors are socially suboptimal, but might seem optimal to people (keeping resources to oneself).
This can be partially attributed to a lack of trading, but also hints at a common human intuition that blocking off regions with resources should be an effective strategy. In contrast, the AI agents learn a strategy that does not include blocking: they might profit from trading and should not waste time on building houses to block off regions.
\item {\em Learning effects}. As human participants experience multiple episodes, their strategy improves, as observed, for example, through lower average productivity and worse social metrics during qualification rounds.
This learning effect is partially controlled for by randomizing the order of tax models presented to participants
and only using experimental episodes
where participants
have already participated in one or more qualification rounds.
\item We set the payoff per house for each human agent using a preset skill.
For real people, this is not the only factor that affects the expected average payoff. For instance, people can have different strategies in the simulation, which affects their average payoff and hence their implied skill. Hence, varying skill and payoff as a simulation setting only partially emulates the effect of skill on the expected payoff and utility that people experience.
\end{itemize}

Considering all  these factors, we find these results for the AI Economist in the presence of human participants  encouraging. The AI-driven tax model did not require knowledge of economic theory, did not require that we estimate the tax elasticity of labor, and was nevertheless able to learn a well-performing tax policy for use with human participants \emph{tabula rasa}. We were able to apply the model without requiring recalibration of tax rates. The only calibration was to scale down the income brackets by a factor of three to adjust for the relative productivity of human and AI agents and enable all income brackets to be exercised.

\medskip

\emph{We emphasize that we do not endorse the particular tax schedule determined by the AI Economist for use in the real economy.}

\medskip

Still, the encouraging transfer performance suggests there is potential for building AI-driven tax models that can find application to the real world, as a new tool to be used by governments. Moreover, given that the AI tax policy, which is dynamic in that its tax schedule changes across tax periods, substantially outperforms the Saez formula in the AI simulations, an interesting direction for future research is to develop experiments that can inform ways with which dynamic tax models can be applied to human settings.

\section{Conclusion}
\label{sec:conclusion}

We believe the intersection of machine learning and economics presents a wide range of exciting research directions, and gives ample opportunity for new machine learning advances that will have significant positive social impact.
Our vision for the AI Economist is to enable an objective study of the impact of economic policies on real-world economies, at a level of complexity that traditional economics research cannot easily address. 

For tax policies in particular, we are hopeful that this kind of research can increase equality and productivity in the real world, helping to promote more just and healthy economies. We also hope that the AI Economist can foster transparency, reproducibility, and open and facts-based discussion about applying machine learning to economic decision-making, through our public research publications and open-source code. As such, we hope that future economic AI models can robustly and transparently augment real-world economic policy-making and in doing so improve social welfare.

In this paper, the economic agents and social planner were trained using model-free RL in AI-based, economic simulations.
A key benefit of using model-free RL is flexibility: for instance, any social objective can be used as the reward function for the planner. Moreover, it does not need any prior world knowledge to find a well-performing tax policy.
However, this approach assumes that the inputs and outputs to the agents' and planner's policy models are sufficient and well-defined. For instance, the planner should be able to observe all state information that is relevant for determining the optimal tax policy. Our initial experiments with human participants suggest that, in our problem setting, the state observed by the planner was sufficient to generalize well to human agents.
In future work, it would be interesting to explore which state information of the real world should be captured by economic simulators to enable generalization of policies from simulation to the real world.

This work is suggestive of the promise of AI-based, economic simulators for  learning economic policies with the potential to transfer well to the real world. 
We demonstrated that economic simulators can yield AI agents with economic behaviors that are consistent with economic intuition, for example agents that specialize as a consequence of their inherent skill level.
Our experiments also show that policies trained in such simulators can transfer well to settings with human participants, albeit for our limited problem setting.

Of course, these kinds of economic simulations still have many limitations. They do not yet model human-behavioral factors and interactions between people, including other-regarding utilities, and consider a relatively small economy.
Moreover, the concept of skill and the associated payoff, as used in our work, is still a limited representation of economic behavior in the real world. For instance, highly-skilled workers are not paid the same hourly wage across different industries, and skill might be hard to clearly define and measure clearly in certain professions.
Future simulations could improve the fidelity of simulated economic behavior by making use of real-world economic data, while advances in large-scale RL and engineering could increase the scope of economic simulations.

\section{Ethics and Normative Aspects}
\label{sec:ethics}
Ethics, trust, and transparency are an integral part of Salesforce’s approach to AI research.
While the current version of the AI Economist provides only a limited representation of the real world, we recognize that it could be possible to manipulate future, large-scale iterations of the AI Economist to increase inequality and hide this action behind the results of an AI system.

Furthermore, either out of ignorance or malice, bad training data may result in biased recommendations, particularly in cases where users will train the tool using their own data. For instance, the exclusion from the model of communities and segments of the work-force that are under-represented in training data might lead to bias in AI-driven tax models. This work also opens up the possibility of using richer, observational data to set individual taxation, an area where we anticipate a strong need for robust debate.

Economic simulation enables studying a wide range of economic incentives and their consequences, including models of stakeholder capitalism.
However, the simulation used in this work is not an actual tool that can be currently used with malintent to reconfigure tax policy.
We encourage anyone utilizing the AI Economist to publish a model card and data sheet that describes the ethical considerations of trained AI-driven tax models to increase transparency, and by extension, trust, in the system.
Furthermore, we believe any future application or policy built on economic simulations should be built on inspectable code and subject to full transparency.

In order to responsibly publish this research, we have taken the following measures:

\begin{itemize}
\item To ensure accountability on our part, we have consulted academic experts on safe release of code and ensured we are in compliance with their guidance. We shared the paper and an assessment of the ethical risks, mitigation strategies, and assessment of safety to publish with the following external reviewers:
Dr. Simon Chesterman, Provost’s Chair and Dean of the National University of Singapore Faculty of Law, and Lofred Madzou, AI Project Lead at the World Economic Forum's Center for the Fourth Industrial Revolution.
None of the reviewers identified additional ethical concerns or mitigation strategies that should be employed. All affirmed that the research is safe to publish.
\item To increase transparency, we are publishing this technical paper, as well as a blog post, thereby allowing robust debate and broad multidisciplinary discussion of our work.
\item To further promote transparency, we will have a timed open-source release of our environment and sample training code for the simulation.
This does not prevent future misuse, but we believe, at the current level of fidelity, transparency is key to promote grounded discussion and future research.
\end{itemize}

With these mitigation strategies and other considerations in place, we believe this research is safe to publish.
Furthermore, this research was not conducted with any corporate or commercial applications in mind.

\bibliographystyle{unsrtnat}
\bibliography{references,references_neurips}

\appendix
\section{Details of Environment}
\label{sec:details_of_environment}

This section provides a more exhaustive description of the environment dynamics and the observations (and, where appropriate, actions) available to agents and the social planner.
We describe these separately for each of the mechanics used to construct the simulation.
Note that observations and actions are essentially the inputs and outputs of the neural network policies trained using RL.
Planner observations/actions are irrelevant for baseline tax models, where tax rates are either fixed or calculated formulaically.

\paragraph{World Dynamics.}
The \textit{Gather-and-Build} environment is organized over a 2D grid.
Grid cells can be occupied by agents, resources, houses, or other landmarks such as water.
At the start of each episode, certain cells are designated as `source' cells that function to spawn new resource units.\footnote{For our purposes, we use the same, fixed layout of source cells each episode.}
A given source cell only spawns a single type of resource, i.e. wood or stone.
If an agent moves to a cell that contains a resource, that resource is added to the agent's inventory and removed from the world at that location.
At the start of each timestep, resources randomly re-spawn at empty source cells according to the regeneration probability.

The state of the world is represented as a $H \times W \times C$ tensor, where $H$ and $W$ are the size of the world and $C$ is the number of unique entities that may occupy a cell, and the value of a given element indicates that a particular entity is occupying the associated location.
The social planner is able to observe the full world state tensor, while agent observations are restricted to views of the state tensor from a narrower, egocentric spatial window.
Our experiments use a world of size 25-by-25, where agent observations have size 11-by-11.
Agent spatial observations are padded as needed when their observation window extends beyond the world grid.

\paragraph{Movement and Gathering.}
Agents must navigate the world in order to collect resources and build new houses.
The action space of the agents includes 4 actions for moving up, down, left, and right.
Agents are restricted from moving on top of water cells, cells occupied by other agents, and cells containing houses built by other agents.
In this way, agents may create difficulty for other agents by restricting their available paths.

Before resources may be sold or used to build houses, they must be collected from the world.
An agent collects resources by moving itself on top of a resource-populated source cell.
By default, this adds a single unit of the collected resources to the agent's inventory, with the possibility of a bonus unit also being collected, the probability of which is determined by the agent's collecting skill.

Agents observe the state of their inventories (including wood, stone, and coin) as well as their own collecting skill.
The planner is also able to observe agents' inventories but cannot observe skill values.

\paragraph{Building.}
When an agent has both stone and wood in its inventory, it may spend one unit of each in order to construct a house.
The action space of the agents includes 1 action for building.
Agents are restricted from building on source cells as well as locations where a house already exists.
Building places a house at the location occupied by the agent and adds coin to the agent's inventory, the amount of which is determined by its building skill.
Agents observe their own building skill, which the planner is not able to observe.

\paragraph{Trading.}
Agents can buy and sell resources from one another through a trading mechanism structured as a \textit{continuous double auction}.
This is conceptually similar to a commodities or stock exchange, where participants do not interface directly but instead submit bids and asks to a market, which identifies and executes valid trades.
The action space of the agents includes 44 actions for trading, representing the combination of 11 price levels ($0, \ldots, 10$ coin), 2 directions (bids and asks), and 2 resources (wood and stone).
Note: agents buy resources by submitting bids and sell resources by submitting asks.
Each trade action therefore maps to a single order (i.e. bid 3 coin for 1 wood, ask for 5 coin in exchange for 1 stone, etc.).
Once an order is submitted, it remains open until either it is matched (in which case a trade occurs) or it expires (after 50 timesteps).
Agents are restricted from having more than 5 open orders for each resource and are restricted from placing orders that they cannot complete: they cannot bid with more coin than they possess and cannot submit asks for resources that they do not have.

A bid/ask pair form a valid trade if they are for the same resource and the bid price matches or exceeds the ask price.
When a new order is received (i.e. bid 3 coin for 1 wood) it is compared against complementary orders to identify potential valid trades.
When a single bid (ask) could be paired with multiple existing asks (bids), priority is giving the ask (bid) with the lowest (highest) price; in the event of ties, priority then is given to the oldest existing order.
Once a match is identified, the trade is executed using the price of whichever order was placed first.
As an example, if the market receives a new bid that offers 8 coin for 1 stone and the market has two open asks offering 1 stone for 3 coin and 1 stone for 7 coin, respectively, the market would pair the bid with the first ask and a trade would be executed for 1 stone at a price of 3 coin:
the bidding agent loses 3 coin and gains 1 stone and the asking agent loses 1 stone and gains 3 coin.
Once a bid and ask are paired and the trade is executed, both orders are removed from the market.

The state of the market is captured by the number of outstanding bids and asks at each price level for each resource.
Agents observe these counts both for their own bids/asks as well as the cumulative bids/asks of other agents (representing the bids/asks that they could respond to).
The planner observes the cumulative bids/asks of all agents.
In addition, both agents and the planner observe some historical information from the market: (for each resource) the average trading price as well as the number of trades at each price level.

\paragraph{Taxation and Redistribution.}
The main text describes the general implementation of periodic, bracketed taxes used in our environment (Section \ref{sect:brackettax}).
On the first timestep of each tax period, the planner sets the marginal tax rates that will be used to collect taxes when the tax period ends.
For baseline models, these are set either formulaically or using fixed rates.
For taxes controlled by a deep neural network (i.e. AI Economist), the action space of the planner is divided into seven action subspaces, one for each tax bracket: $\{0, 0.05, 0.10, \ldots, 1.0\}^7$.
Each subspace denotes the set of discretized marginal tax rates that the planner may select.\footnote{Discretization of tax rates only applies to deep learning networks.}
The action space takes this form because, when setting new rates, the planner samples 7 rates at once.

Agents observe:
\begin{itemize}
    \item The tax rates of the current tax period
    \item The marginal rate at the income level earned within the current period so far
    \item Indicators of the temporal progress of the current tax period
    \item The set of sorted and anonymized incomes the agents reported in the previous period
\end{itemize}
The planner observes the same information as well as the non-anonymized income and marginal tax rate (at that income) of each agent in the previous period.

\paragraph{Action Details.}
In our implementation, both agents and planners use discrete action spaces.
One advantage of this choice is that it grants easier control over which actions the agent/planner models can sample at a given time.
We encode this using `action masks' which we use to ensure that the policy network assigns effectively 0 probability to restricted actions.
Action masking is useful for preventing invalid actions and is how we control the tax annealing (see Section \ref{sec:trainingstrategy}) for the AI Economist experiments.

In addition to the actions described above, we include a \texttt{NO-OP} action (``no operation'') in each action space.
(For the planner, each of the 7 action subspaces includes a \texttt{NO-OP} action.)
The \texttt{NO-OP} action is interpreted as essentially taking no action, allowing the agent to ``idle'' and the planner to leave a bracket's tax rates unchanged between periods.

Most importantly, we use these implementation features to enable the planner to observe every timestep while only acting at the start of each new tax period.
For timesteps other than those at the start of a tax period we simply use the action mask to enforce that only \texttt{NO-OP} actions are sampled.
This allows the planner to use rich temporal information while also ensuring that policy gradients are only propagated from the action samples used to control taxes.

\section{Training Hyperparameters and Experiment Settings}
\label{sec:hyperparameters}

\begin{table}[ht!]
    \begin{small}
        \begin{center}
            \begin{tabular}[c]{lll}
                Parameter & & Value \\
                \hline
                \hline

                Training algorithm & & \textit{ppo} \\
                Number of parallel environment replicas & & 60 \\
                Sampling horizon (steps per replica) & $\samplinghorizon$ & 200 \\
                SGD minibatch size & & 3000 \\
                SGD sequence length & & 50 \\
                Policy updates per horizon (agent) & & 16 \\
                Policy updates per horizon (planner) & & 4 \\

                CPUs & & 15 \\
                GPUs & & 2 \\

                \hline

                Learning rate (agent) & & 0.0003 \\
                Learning rate (planner) & & 0.0001 \\
                Entropy regularization coefficient (agent) & & 0.025 \\
                Entropy regularization coefficient (planner) & & 0.1 \\
                Gamma & $\df$ & 0.998 \\
                GAE lambda & & 0.98 \\
                Gradient clipping & & 10 \\
                Value function loss coefficient & & 0.05 \\

                \hline

                Number of convolutional layers & & 2 \\
                Number of fully-connected layers & & 2 \\
                Fully-connected layer dimension (agent) & & 128 \\
                Fully-connected layer dimension (planner) & & 256 \\
                LSTM cell size (agent) & & 128 \\
                LSTM cell size (planner) & & 256 \\
                All agents share weights & & True \\
                Value/Policy networks share weights & & False \\

                \hline

                Planner gets spatial info & & True \\
                Agents get full spatial observation & & False \\
                Agent spatial observation box half-width & & 5 \\

                \hline

                Phase \textit{one} training duration & & 50M steps \\
                Phase \textit{two} training duration & & 400M steps \\
                Phase \textit{two} initial max $\mtaxrate$ & & 10\% \\
                Phase \textit{two} tax annealing duration & & 54M steps \\

            \end{tabular}
        \end{center}
        \caption{Training hyperparameters.}
        \label{tab:training_settings_ai}
    \end{small}
\end{table}
For each experiment, experience collection was parallelized over 60 replicas of the environment.
Each training iteration involved collecting $\samplinghorizon$ steps from each replica (using the latest policy parameters), followed by a round of parameter updates using the collected samples.
With a sampling horizon of 200 timesteps and 60 environment replicas, a total of 12000 timesteps were sampled per training iteration.
Since each ``agent'' experiences its transition per timestep, this is actually a total of 60000 transitions: 48000 for the 4 agents and 12000 for the planner.
When doing policy updates, we divided each such set of transitions into minibatches of size 3000 and perform one gradient update per minibatch.
Therefore, each training iteration involved 16 updates to the agent parameters and 4 to the planner parameters.
We use PPO to accomodate multiple updates per training iteration.

Tables \ref{tab:training_settings_ai} and \ref{tab:env_settings_ai} provide details regarding the training hyperparameters and environment settings, respectively, used in our AI experiments.
\begin{table}[ht!]
    \begin{small}
        \begin{center}
            \begin{tabular}[c]{lll}
                Parameter & & Value \\
                \hline
                \hline

                Number of agents & $N$ & 4 \\
                Episode length & $\eplen$ & 1000 \\
                World height & & 25 \\
                World width & & 25 \\
                Resource respawn probability & & 0.01 \\
                Max resource health & & 1 \\
                Skill distribution & & \textit{pareto} \\
                Starting agent coin & $\money_{i, 0} $ & 0 \\

                \hline

                Iso-elastic utility exponent & $\eta$ & 0.23 \\
                Move labor & & 0.21 \\
                Gather labor & & 0.21 \\
                Trade labor & & 0.05 \\
                Build labor & & 2.1 \\

                \hline

                Minimum build payout & & 10 \\
                Build payment max skill multiplier & & 3 \\
                House lifetime & & $\inf$ \\

                \hline

                Max bid/ask price & & 10 \\
                Max bid/ask order duration & & 50 \\
                Max number of open orders per resource & & 5 \\

                \hline

                Tax period duration & $\taxperiodlen$ & 100 \\
                Min bracket rate & & 0\% \\
                Max bracket rate & & 100\% \\
                Rate discretization (AI Economist) & & 5\% \\
                Bracket cutoffs & $\{\bracketcutoff_0, \ldots, \bracketcutoff_B\}$ & \textit{us-federal} \\
                social welfare weights (Saez formula) & $g_i$ & \textit{inverse-income} \\

                \end{tabular}
        \end{center}
        \caption{Environment settings.}
        \label{tab:env_settings_ai}
    \end{small}
\end{table}

\section{Details on Experiments with Human Participants}
\label{app:human_exp_details}
\begin{figure}[t!]
    \begin{center}
        \begin{small}
            \begin{center}
                \includegraphics[width=0.4\linewidth]{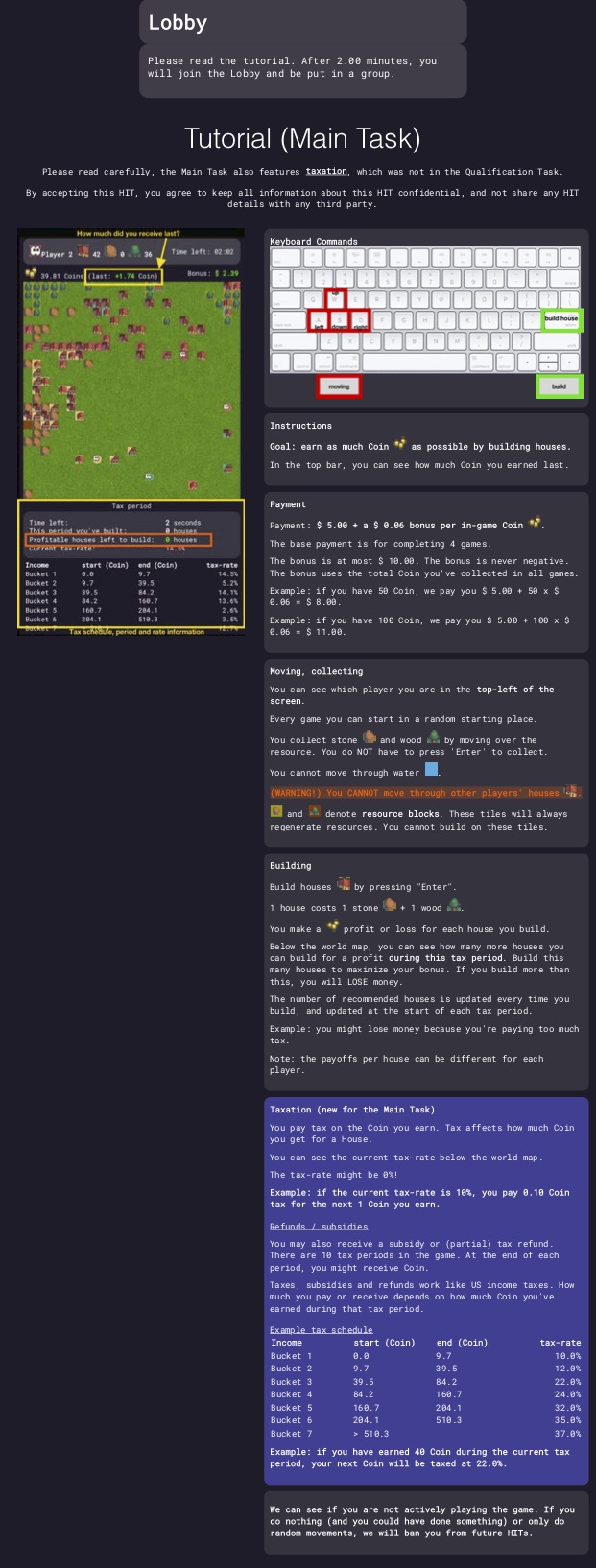}
            \end{center}
            \caption{The starting point for each experiment was the lobby and tutorial (Figure \ref{fig:human_exp_gui_tutorial}). The tutorial explained all rules of the world and the objective for each participant. It also explained how the variable part of payment for the experiment was computed. In addition, an example of the graphical interface and keyboard controls was shown. Participants were warned not to idle in the game. Participants were given 2 minutes before the start of the experiment to read the tutorial. After this initial period, as soon as there were enough participants to form a group, the lobby system presented the participants in the new group with the experiment's main page.}
            \label{fig:human_exp_gui_tutorial}
        \end{small}
    \end{center}
\end{figure}
\begin{figure}[t!]
    \begin{center}
        \begin{small}
            \begin{center}
                \includegraphics[width=0.5\linewidth]{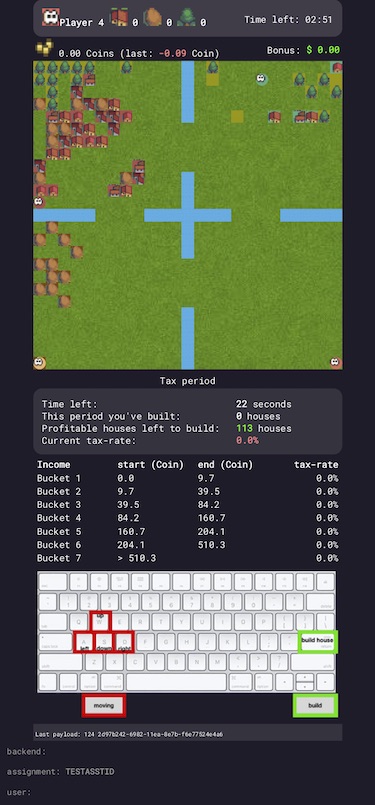}
            \end{center}
            \caption{The main experiment's graphical user interface (Figure \ref{fig:human_exp_gui_main}) showed (from top to bottom): the endowment of the agent, the remaining time in the episode, the bonus amount earned so far, the spatial state of the world, the tax information and current tax rate, the time left, the number of houses built and the number of profitable houses left to build. Below all data, a reminder of the controls was shown as well. After an episode was over, participants were returned to the lobby (if the participant had seen less than 4 episodes) or survey (if 4 episodes had been seen).
            }
            \label{fig:human_exp_gui_main}
        \end{small}
    \end{center}
\end{figure}
\begin{figure}[t!]
    \begin{center}
        \begin{small}
            \begin{center}
                \includegraphics[width=0.8\linewidth]{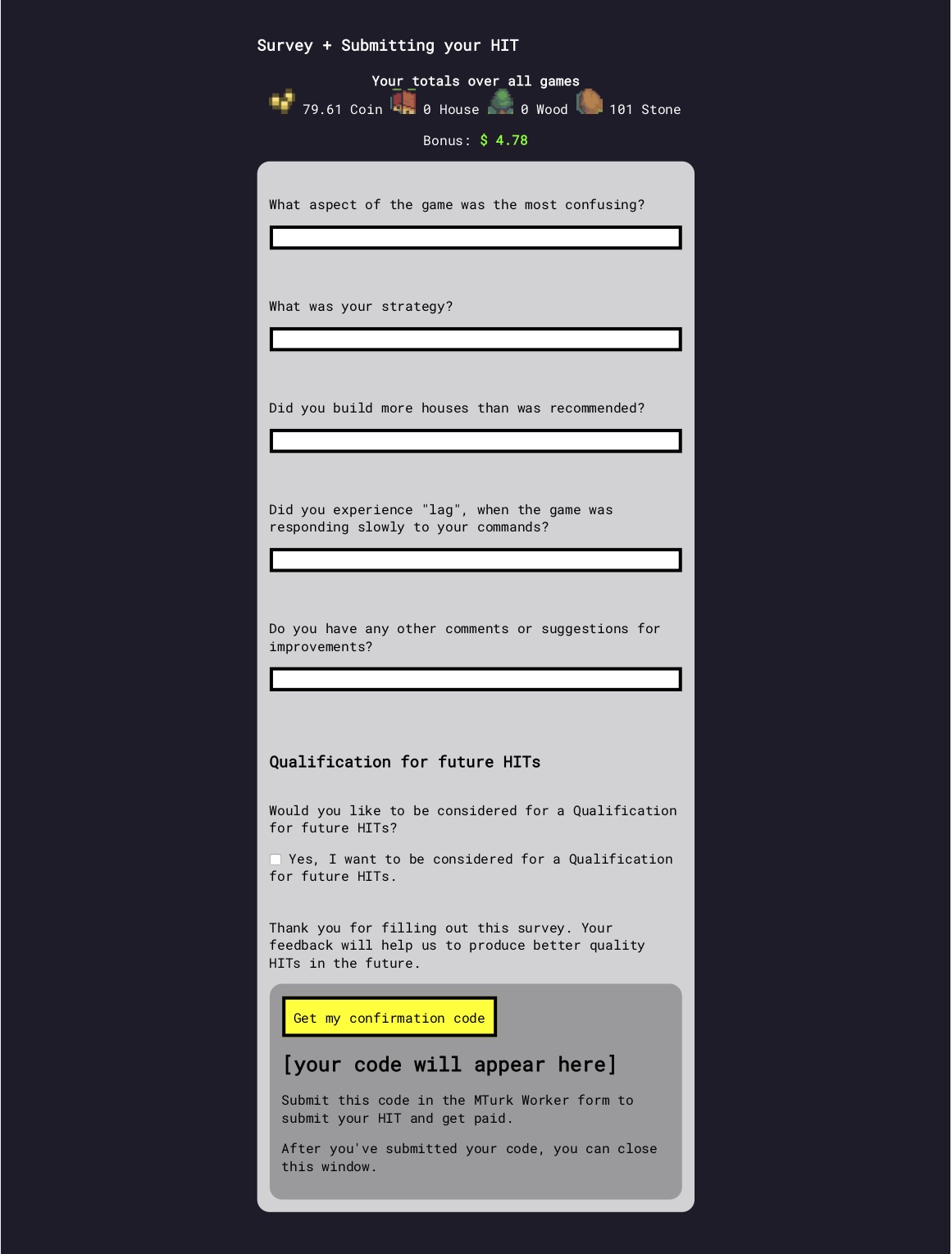}
            \end{center}
            \caption{The post-experiment survey showed how much the participant had collected in terms of resources, coin and bonus. It also asked questions about the participants experience, strategy and general feedback on the experiment. For instance, participants could communicate technical issues, such as lag. At the end of the survey, participants were given a confirmation code that allowed them to confirm successful completion of the task on the Amazon Mechanical Turk platform.}
            \label{fig:human_exp_gui_survey}
        \end{small}
    \end{center}
\end{figure}

All experiment modules used with human participants are shown and described in Figures \ref{fig:human_exp_gui_tutorial} (lobby, tutorial), \ref{fig:human_exp_gui_main} (main graphical interface), and \ref{fig:human_exp_gui_survey} (survey).

\end{document}